\documentstyle[12pt,epsf]{article} \hoffset -0.5in \textwidth 6.00in
\textheight
8.5in  \parskip 7pt \openup1.5\jot \parindent=0.5in
\newskip\humongous \humongous=0pt plus 1000pt minus 1000pt
\def\caja{\mathsurround=0pt}
\def\eqalign#1{\,\vcenter{\openup1\jot \caja
        \ialign{\strut \hfil$\displaystyle{##}$&$
        \displaystyle{{}##}$\hfil\crcr#1\crcr}}\,}
\newif\ifdtup

\def\eqright #1\cr{\noalign{\hfill$\displaystyle{{}#1}$}}
\def\eqleft #1\cr{\noalign{\noindent$\displaystyle{{}#1}$\hfill}}

\def\oldreffmt#1{\rlap{[#1]} \hbox to 2\parindent{}}

\def\figfmt#1{\rlap{Figure {#1}} \hbox to 1in{}}

\def\sectioneq{\def\theequation{\thesection.\arabic{equation}}{\let
\holdsection=\section\def\section{\setcounter{equation}{0}\holdsection}}}%

\def\auto{\eqno(\refstepcounter{equation}\theequation)}
\def\begineq #1\endeq{$$ \refstepcounter{equation}\eqalign{#1}\eqno
	(\theequation) $$}
\def\contlimit{\,{\hbox{$\longrightarrow$}\kern-1.8em\lower1ex
\hbox{${\scriptstyle (a\rightarrow0)}$}}\,}
\def\centeron#1#2{{\setbox0=\hbox{#1}\setbox1=\hbox{#2}\ifdim
\wd1>\wd0\kern.5\wd1\kern-.5\wd0\fi
\copy0\kern-.5\wd0\kern-.5\wd1\copy1\ifdim\wd0>\wd1
\kern.5\wd0\kern-.5\wd1\fi}}
\def\centerover#1#2{\centeron{#1}{\setbox0=\hbox{#1}\setbox
1=\hbox{#2}\raise\ht0\hbox{\raise\dp1\hbox{\copy1}}}}
\def\centerunder#1#2{\centeron{#1}{\setbox0=\hbox{#1}\setbox
1=\hbox{#2}\lower\dp0\hbox{\lower\ht1\hbox{\copy1}}}}
\def\lsim{\;\centeron{\raise.35ex\hbox{$<$}}{\lower.65ex\hbox
{$\sim$}}\;}
\def\gsim{\;\centeron{\raise.35ex\hbox{$>$}}{\lower.65ex\hbox
{$\sim$}}\;}

\def\super#1{\ifmmode \hbox{\textsuper{#1}}\else\textsuper{#1}\fi}
\def\textsuper#1{\newcount\holdspacefactor\holdspacefactor=\spacefactor
$^{#1}$\spacefactor=\holdspacefactor}

\def\getcite#1,{\advance\citenumber by1
\def\getcitearg{#1}\def\lastarg{@}
\ifnum\citenumber=1
\ref{#1}\let\next=\getcite\else\ifx\getcitearg\lastarg\let\next=\relax
\else ,\ref{#1}\let\next=\getcite\fi\fi\next}

\def\pom{{\rm P\kern -0.53em\llap I\,}}
\def\spom{{\rm P\kern -0.36em\llap \small I\,}}
\def\sspom{{\rm P\kern -0.33em\llap \footnotesize I\,}}

\relax

\newskip\humongous \humongous=0pt plus 1000pt minus 1000pt
\def\caja{\mathsurround=0pt}
\def\eqalign#1{\,\vcenter{\openup1\jot \caja
        \ialign{\strut \hfil$\displaystyle{##}$&$
        \displaystyle{{}##}$\hfil\crcr#1\crcr}}\,}
\newif\ifdtup

\def\eqright #1\cr{\noalign{\hfill$\displaystyle{{}#1}$}}
\def\eqleft #1\cr{\noalign{\noindent$\displaystyle{{}#1}$\hfill}}

\def\oldreffmt#1{\rlap{[#1]} \hbox to 2\parindent{}}

\def\figfmt#1{\rlap{Figure {#1}} \hbox to 1in{}}

\def\auto{\eqno(\refstepcounter{equation}\theequation)}
\def\begineq #1\endeq{$$ \refstepcounter{equation}\eqalign{#1}\eqno
	(\theequation) $$}
\def\contlimit{\,{\hbox{$\longrightarrow$}\kern-1.8em\lower1ex
\hbox{${\scriptstyle (a\rightarrow0)}$}}\,}
\def\centeron#1#2{{\setbox0=\hbox{#1}\setbox1=\hbox{#2}\ifdim
\wd1>\wd0\kern.5\wd1\kern-.5\wd0\fi
\copy0\kern-.5\wd0\kern-.5\wd1\copy1\ifdim\wd0>\wd1
\kern.5\wd0\kern-.5\wd1\fi}}
\def\centerover#1#2{\centeron{#1}{\setbox0=\hbox{#1}\setbox
1=\hbox{#2}\raise\ht0\hbox{\raise\dp1\hbox{\copy1}}}}
\def\centerunder#1#2{\centeron{#1}{\setbox0=\hbox{#1}\setbox
1=\hbox{#2}\lower\dp0\hbox{\lower\ht1\hbox{\copy1}}}}
\def\lsim{\;\centeron{\raise.35ex\hbox{$<$}}{\lower.65ex\hbox
{$\sim$}}\;}
\def\gsim{\;\centeron{\raise.35ex\hbox{$>$}}{\lower.65ex\hbox
{$\sim$}}\;}

\def\super#1{\ifmmode \hbox{\textsuper{#1}}\else\textsuper{#1}\fi}
\def\textsuper#1{\newcount\holdspacefactor\holdspacefactor=\spacefactor
$^{#1}$\spacefactor=\holdspacefactor}

\def\getcite#1,{\advance\citenumber by1
\ifnum\citenumber=1
\ref{#1}\let\next=\getcite\else\ifx#1@\let\next=\relax
\else ,\ref{#1}\let\next=\getcite\fi\fi\next}

\def\upon #1/#2 {{\textstyle{#1\over #2}}}
\relax

\def\mainhead#1{\setcounter{equation}{0}\addtocounter{section}{1}
  \vbox{\begin{center}\large\bf #1\end{center}}\nobreak\par}
\sectioneq

\def\til#1{\centeron{\hbox{$#1$}}{\lower 2ex\hbox{$\char'176$}}}
\def\tild#1{\centeron{\hbox{$\,#1$}}{\lower 2.5ex\hbox{$\char'176$}}}
\def\sumtil{\centeron{\hbox{$\displaystyle\sum$}}{\lower
-1.5ex\hbox{$\widetilde{\phantom{xx}}$}}}

\def\pom{{\rm P\kern -0.53em\llap I\,}}
\def\spom{{\rm P\kern -0.36em\llap \small I\,}}
\def\sspom{{\rm P\kern -0.33em\llap \footnotesize I\,}}

\newcommand{\bit}{\begin{itemize}}
\newcommand{\eit}{\end{itemize}}

\newcommand{\beq}{\begin{equation}}
\newcommand{\eeq}{\end{equation}}
\newcommand{\beqa}{\begin{eqnarray}}
\newcommand{\eeqa}{\end{eqnarray}}

\begin{document}

\begin{titlepage}
\rightline{\vbox{\halign{&#\hfil\cr
&ANL-HEP-PR-95-12\cr }}}

{}~
\vspace{1in}

\begin{center}

{\bf PROPERTIES OF THE SCALE INVARIANT $O(g^4)$ LIPATOV KERNEL}
\medskip

Claudio Corian\`{o} and Alan R. White
\footnote{Work supported by the U.S. Department of Energy, Division of High
Energy Physics, Contract\newline W-31-109-ENG-38}
\\ \smallskip
High Energy Physics Division, \\
Argonne National Laboratory, \\
Argonne, IL 60439. \\

\end{center}

\begin{abstract}

We study the scale-invariant $O(g^4)$ kernel which appears as an infra-red
contribution in the BFKL evolution equation and is constructed via
multiparticle $t$-channel unitarity. We detail the variety of Ward identity
constraints and infra-red cancellations that characterize its infrared
behaviour. We give an analytic form for the full non-forward kernel. For the
forward kernel controlling parton evolution at small-x, we give an impact
parameter representation, derive the eigenvalue spectrum, and demonstrate a
holomorphic factorisation property related to conformal invariance. The results
show that, at next-to-leading-order, the transverse momentum infra-red
region may produce a strong reduction of the BFKL small-x behavior.

\end{abstract}

\end{titlepage}

\mainhead{1. INTRODUCTION}

The BFKL equation\cite{bfkl} describes the leading log small-x evolution of
parton distributions in QCD. In recent papers\cite{ker,uni} we have proposed
that a scale-invariant approximation to the next-to-leading order kernel can
be directly constructed from $t$-channel unitarity. We have also
presented\cite{spec} results on the structure and eigenvalue spectrum of the
new, $O(g^4)$, kernel. Our purpose in this paper is to derive various
properties of this kernel which have been utilised in \cite{spec}, to elaborate
on derivations outlined there, and also to describe additional new results.

In a companion paper\cite{exp} we give a full discussion of the $t$-channel
unitarity derivation of reggeon kernels which was only sketched in
\cite{uni}. We show how the angular momentum plane structure of a gauge
theory can be obtained by expanding around $j=1$ (the analogue of expanding in
powers of logarithms in momentum space) in combination with the Ward identity
constraints and group structure that define the theory. At $O(g^4)$, the
four-particle nonsense states provide an unambiguous infra-red contribution
which can be written as the sum of scale-invariant transverse momentum
integrals that we study in this paper. Here we will not discuss either this
derivation or the very important issue of how the relevant physical scales
should be introduced. Although this is crucial for determining physical
contributions correctly, we believe the mathematical properties of the new
kernel are of sufficient interest in their own right.

We will particularly emphasize those properties that the new kernel shares
with the $O(g^2)$ kernel. It is, of course, scale-invariant, it is also
infra-red finite, both before integration and as an integral kernel, and
satifies Ward Identity constraints at zero momentum transfer. In addition we
will show that there is a component of the parton evolution kernel (that is
the forward kernel) which is separately infra-red finite and whose
eigenvalue spectrum shares many properties of the spectrum of the $O(g^2)$
kernel. In particular we are able demonstrate the important property of
holomorphic factorization for the eigenvalues which is closely related to
conformal symmetry in the conjugate impact parameter space\cite{kir}.

We begin by reviewing the BFKL kernel in Section 2. This serves to introduce
language and to focus on those properties of the leading-order kernel which
we generalize. In Section 3 we present the $O(g^4)$ kernel and elaborate on
the variety of Ward identity properties and infra-red cancellations. Section
4 is devoted to the evaluation of the two-dimensional box diagram as a sum
of logarithms. This allows us to give an analytic expression for the
complete kernel. The remaining Sections are devoted to the forward kernel.
In Section 5 we discuss the structure of this kernel and also introduce a
dimensionally regularized form to prepare for the eigenvalue evaluation.
Section 6 contains expressions in impact parameter space for the distinct
components of the forward kernel. In Section 7 we compute the spectrum of
both the four-particle kernel and the kernel introduced in \cite{spec} which
contains also two-particle nonsense states. We show that a large
reduction of the BFKL power growth of parton distributions at small-x can
occur. Finally we briefly discuss the significance of our results in a
concluding Section.

\mainhead{2. THE $O(g^2)$ PARTON AND REGGEON KERNELS}

In this Section we review properties of the BFKL equation and
kernel\cite{bfkl,lip}. This will introduce our notation and establish various
properties which we want to compare with for the $O(g^4)$ kernel.

The most familiar application of the BFKL equation is as an evolution
equation for parton distributions at small-x i.e.
$$
\eqalign{ {\partial \over \partial (ln {1 / x})}F(x,k^2) ~=~\tilde{F}(x,k^2)~+~
{1 \over (2\pi)^3}\int {d^2k' \over (k')^4} ~K(k,k') F(x,{k'}^2) }
\auto\label{eve}
$$
where, if the gauge group is SU(N), the ``parton'' kernel $K(k,k')$ is given by
$$
\eqalign{  (Ng^2)^{-1}K(k,q)~=~\Biggl(-~\delta^2(k-k')k^6
\int {d^2p \over p^2(k-p)^2}~+~{2k^2{k'}^2 \over (k-k')^2}~\Biggr)}
\auto
$$

This equation originates, however, in Regge limit calculations\cite{bfkl} where
also a non-forward (i.e.$q \neq 0$ in the following) version is derived. If we
transform to $\omega$ - space (where $\omega$ is conjugate to $ln~{1 \over
x}$), we can write the full non-forward equation in the form
$$
\eqalign{
\omega F(\omega,k,q-k) ~=~\tilde{F}~+~ {1 \over 16\pi^3}\int {d^2k' \over
(k')^2(k'-q)^2}~K(k,k',q) F(\omega,k',q-k') }
\auto\label{ome}
$$
where the ``reggeon'' kernel $K(k,k',q)~=~K^{(2)}_{2,2}(q-k,k,k',q-k')$ now
contains three kinematic forms. (Throughout this paper we normalise our
transverse momentum integrals by $(16\pi^3)^{-1}$, rather than $(2\pi)^{-3}$
as in our previous papers.) We write, therefore,
$$
\eqalign{ {1 \over Ng^2}~K^{(2)}_{2,2}(k_1,k_2,k_3,k_4)~
 =~\sum \Biggl(&-{1 \over
2}k_1^4J_1(k_1^2)k_2^2(16\pi^3)\delta^2(k_2-k_3)\cr
&+~{k_1^2k_3^2 \over (k_1-k_4)^2}
{}~~-~~{1 \over 2}(k_1+k_2)^2\Biggr)\cr
&\equiv~~K^{(2)}_1~+~K^{(2)}_2~+K^{(2)}_3 ~~.}
\auto\label{2,2}
$$
where
$$
J_1(k^2)~=~{1 \over 16\pi^3}\int {d^2k' \over
(k')^2(k'-k)^2}
\auto\label{j1}
$$
and the $\sum$ implies that we sum over permutations of both the initial and
the final state. That is, we add to the explicit expressions we have given, the
same expressions with the suffices $1$ and $2$ interchanged and then add
further expressions with $3$ and $4$ interchanged. (In previous papers we
have used a notation involving summation over permutation of $1$ and $2$
only. The present notation is consistent with the diagrammatic notation we
use.)

In the following we will utilise transverse momentum diagrams extensively.
We construct these diagrams using the components illustrated in Fig.~2.1.

\vspace{.2in}

\epsffile{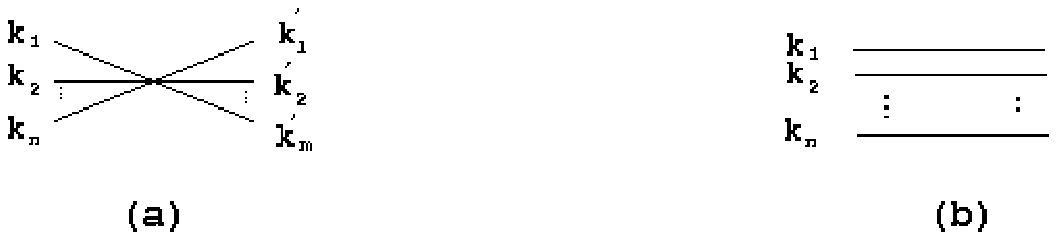}
\begin{center}
Fig.~2.1 (a)vertices and (b) intermediate states forming transverse momentum
diagrams
\end{center}

\noindent For each vertex, illustrated in Fig.~2.1(a), we write a factor
$$
16\pi^3\delta^2(\sum k_i~  - \sum k_i')(\sum k_i~)^2
$$
For each intermediate state, illustrated in Fig.~2.1(b), we write a factor
$$
(16\pi^3)^{-n}\int d^2k_1...d^2k_n~ /~k_1^2...k_n^2
$$
We will define dimensionless kernels and components of kernels by
including a factor $16\pi^3\delta^2(\sum k_i - \sum {k'}_i)$
in their definition. We denote this by a hat e.g.
$$
\hat{K}^{(2)}_{2,2}(k_1,k_2,k_3,k_4)~=~
16\pi^3\delta^2(k_1+k_2-k_3-k_4) K^{(2)}_{2,2}(k_1,k_2,k_3,k_4)~
$$
(In \cite{spec} we did not use the hat but instead used a tilde to denote a
kernel without the $\delta$-function.) This defines kernels that are formally
scale-invariant (even though potentially infra-red divergent). The
diagrammatic representation of $\hat{K}^{(2)}_{2,2}$ is then as shown in
Fig.~2.2,

\vspace{.2in}

\epsffile{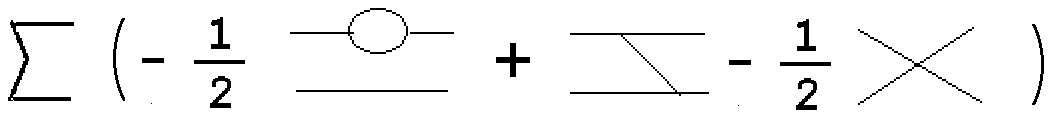}
\begin{center}
Fig.~2.2 Diagrammatic representation of $\hat{K}^{(2)}_{2,2}$
\end{center}

\noindent where the summation sign again implies a sum over all permutations
of the initial and final momenta.

There are two crucial properties of ${K}^{(2)}_{2,2}$ that we will
generalise in the following which are direct consequences of the gauge
invariance of the theory.

\begin{itemize}

\item Ward identity constraints\cite{ker,exp} are satisfied i.e.
$$
\eqalign{ K^{(2)}_{2,2}(k_1,k_2,k_3,k_4) ~~\to ~~0~~, k_i~\to~ 0~, i=~1,..,4}
\auto\label{war}
$$

\item Infra-red finiteness as an integral kernel i.e.
$$
\eqalign{ \int {d^2k_1 \over k_1^2} {d^2k_2 \over k_2^2} \delta^2(q-k_1-k_2)
K^{(2)}_{2,2}(k_1,k_2,k_3,k_4) ~~~~~~is ~~finite}
\auto\label{fin}
$$

\end{itemize}

\noindent These two properties actually determine the relative magnitude of
the three kinematic forms $K^{(2)}_1, K^{(2)}_2,$ and $K^{(2)}_3$. It is very
simple to demonstrate them diagrammatically. First we note two simple rules
for obtaining the $k_i \to 0$ limit for any transverse momentum diagram.

\begin{itemize}

\item $k_i \to 0$ gives zero if the line carrying $k_i$ is the single line of a
1-2,  2-1, or 1-1 vertex.

\item In general, $k_i \to 0$ gives the subdiagram obtained by removing the
line carrying $k_i$.

\end{itemize}

As we outlined in \cite{ker} and discuss further in \cite{exp}, the Ward
identity constraint (\ref{war}) should be satisfied by any reggeon amplitude
in a gauge theory. It is easily proved diagrammatically. Using a dotted line
to denote the $k_i$ line, we obtain the result shown in Fig.~2.3

\vspace{.2in}

\epsffile{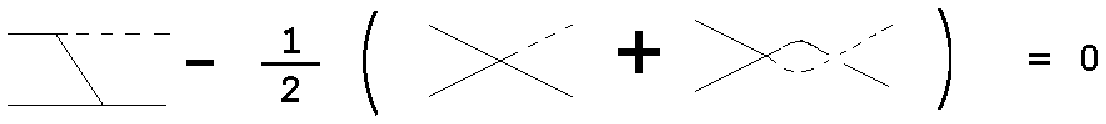}
\begin{center}
Fig.~2.3 The Ward identity constraint for $K^2$
\end{center}

There is a subtlety in applying the first rule above. While the
vertex involved carries a factor of $k_i^2$, if it is attached to a loop
then the loop integral will have a threshold divergence. This divergence
will produce an additional factor of $k_i^{-2}ln[k_i^2]$, apparently nullifying
the Ward identity zero. However, if we regulate the loop integrals the
vertex zero does give the requisite vanishing of the regulated integral. The
cancellation of divergences then leads to the persistence of the zero in the
unregulated kernel. We shall discuss this effect in the $O(g^4)$ kernel in
the next Section. (The trajectory function appearing in $K^{(2)}_1$ is
actually a prime example of this combination of a zero and a divergence. The
effective persistence of the zero is then equivalent to the persistence of
reggeization, i.e. that the gluon Regge trajectory passes through $\omega=0$
at $k^2=0$.)

Note that the Ward identity actually determines the 2-2 vertex appearing in
$K^{(2)}_3$ in terms of the 1-2 coupling appearing in $K^{(2)}_2$. In this way
gauge invariance determines that the complete kernel is written in terms of
a single coupling $g$, which can be identified with the gauge coupling, but
in this context is more correctly identified as a three-reggeon coupling.

We make very little reference to color structure in this paper because we
are discussing only physical kernels that carry zero ($t$-channel) color.
In the $O(g^2)$ kernel the only remnant of the gauge group is then the
overall normalization factor of $N$. The infra-red finiteness property is
actually a zero color property. To demonstrate (\ref{fin}) diagrammatically
we first note that infra-red divergences occur when the momentum $k_i$ of an
internal line vanishes. If we use a mass regulation then, as $m^2~\to ~0$,
this gives
$$
\int {d^2k_i ~f(k_i) / (k_i^2 + m^2)} ~\to~{1 \over 2} \int {dk_i^2
\over (k_i^2+m^2)}
\int_0^{2\pi} d\theta  ~f(0) ~\to~\pi~log[m^2]~f(0)
\auto\label{ird}
$$
where (apart from a factor of $(16\pi^3)^{-1}$) $f(0)$ is obtained from the
original diagram by removing the line carrying $k_i$.

The Ward identity constraint already determines that there is no
divergence in (\ref{fin}) as $k_1 \to 0$ or $k_2 \to 0$ (for non-zero
q). Potential divergences are therefore at $k_{1,2}~=~k_{3,4}$. The
cancellation of infra-red divergences of this kind is demonstrated
diagrammatically in Fig.~2.4.

\medskip

\epsffile{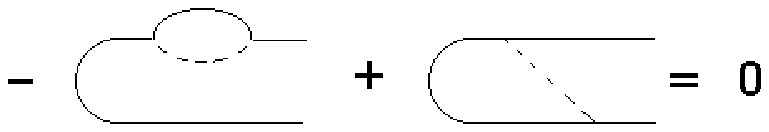}
\begin{center}
Fig.~2.4 Cancellation of infra-red divergences for $K^2$.
\end{center}

To find the $\omega$-plane singularities of $F(\omega,k,q-k)$ we project
(\ref{ome}) on the complete set of orthogonal eigenfunctions
$$
\eqalign{ \phi_{\mu,n}(k)~=~({k}^2)^{\mu}~e^{i n \theta}}~~~~~~~~\mu=~{1
\over2}
+i\nu,~~ n=0,\pm1,\pm2,...
\auto\label{eif}
$$
where $k=(|k|cos\theta,|k|sin\theta)$. (Our definition of the kernel
requires that we keep a factor of $k^{-2}$ in the measure of the
completeness relation for eigenfunctions relative to \cite{lip}).

Note that when $k^2~\sim~{k'}^2~\to~\infty$, $K(k,k',q)~\to K(k,k')$. Since
this region contributes dominantly to the eigenvalue spectrum, the eigenvalues
are independent of $q^2$ and take the form
$$
{Ng^2 \over 2\pi^2} ~\chi(\nu,n)
\auto
$$
where
$$
\chi(\nu,n)~= ~\psi (1) - Re \psi ({|n|+1 \over 2} +i\nu)
{}~~~~~~~\nu ~\epsilon~(-\infty, \infty),~~n=0,\pm 1, \pm2, ...
\auto\label{ei2}
$$
with $\psi (x)~=~{d \over dx}\Gamma(x)$. With
$\tilde{F}~=~\delta^2(k-\tilde{k})$, the solution of (\ref{ome}) at
$q=0$ is the reggeon Green function
$$
F(k,\tilde{k})~=~\sum_{n=-\infty}^{\infty}{e^{in(\theta-\tilde{\theta})}
\over 4\pi^2}
\int_{-\infty}^{\infty} { d\nu~~\bigl(k^2/\tilde{k}^2\bigr)^{i\nu}
\over \omega ~-~{Ng^2 \over 2\pi^2}~\chi(\nu,n)}
\auto\label{gre}
$$
The leading singularity in the $\omega$-plane is given by
$$
\alpha_0 -1 = {Ng^2 \over 2\pi ^2}~\chi(0,0)
\auto\label{alp}
$$
where $\chi(0,0)$ is the leading eigenvalue. If we write $\alpha_s~=~g^2/4\pi$
then since
$$
\chi(0,0)~=~ 2 ln2
\auto\label{chi}
$$
we obtain from (\ref{alp}), the familiar result,
$$
\alpha_0 -1= ({12\alpha_s \over \pi}) ln 2 ~\sim~- {1 \over 2}
\auto
$$
for the small-x power behaviour of parton distributions.

To discuss the conformal symmetry properties of the eigenvalue spectrum
(\ref{ei2}) we rewrite (\ref{ei2}) as\cite{kir}
$$
4\chi(\nu,n)~=~ 4\psi (1) - \psi (m) - \psi (1-m) - \psi (\tilde{m})
- \psi (1 - \tilde{m})
\auto\label{ei3}
$$
where now
$$
m~=~{1 \over 2}~+~i\nu~+~{n \over 2}~,
{}~\tilde{m}~=~{1 \over 2}~+~i\nu~-~{n \over 2}~,
\auto
$$
are conformal weights. $m(1-m)$ and $\tilde{m}(1-\tilde{m})$ are,
respectively, the eigenvalues of the holomorphic and antiholomorphic Casimir
operator of linear conformal transformations.

We can rewrite (\ref{ei3}) as
$$
4\chi(\nu,n)~=~ {\cal F}\bigl[m(1-m)\bigr]~+
{}~{\cal F}\bigl[\tilde{m}(1-\tilde{m})\bigr]
\auto\label{ei4}
$$
where using
$$
\psi(x)~=~\psi(1)~-~\sum_{r=0}^{\infty}\biggl({1 \over r+x}-
{1 \over r+1}\biggr)
\auto
$$
we find\cite{kir}
$$
{\cal F}\bigl[x \bigr]~=~\sum_{r=0}^{\infty}\biggl({2r+1 \over r(r+1)~+~x}-
{2 \over r+1}\biggr)
\auto\label{conf}
$$
That the eigenvalues can be written as a function of $m(1-m)$ plus a
function of $\tilde{m}(1-\tilde{m})$ gives the holomorphic factorization
of the kernel and the conformal symmetry of the BFKL equation\cite{kir}.

\mainhead{3. THE $O(g^4)$ REGGEON KERNEL}

We consider specifically the contribution of ($t$-channel) four-particle
nonsense states to the $O(g^4)$ kernel. In Refs. \cite{uni,spec} we also
considered the contribution obtained by iteration of the two-particle nonsense
states. We will postpone discussion of this component until later Sections.
It is the the interesting mathematical properties of the four-particle
component with which we shall be mostly concerned in this paper. We study,
therefore, the $O(g^4)$ kernel $K^{(4)}_{2,2}$ defined by the sum of
transverse momentum integrals
$$
\eqalign{{1 \over (g^2N)^2} K^{(4n)}_{2,2}(k_1&,k_2,k_3,k_4)
{}~=~K^{(4)}_0~+~K^{(4)}_1~+~K^{(4)}_2~+~K^{(4)}_3~+K^{(4)}_4~}.
\auto\label{sum}
$$
with
$$
\eqalign{K^{(4)}_0~=~{1 \over 2} ~
\sum ~ k_1^4k_2^4J_1(k_1^2)J_1(k_2^2)(16\pi^3)\delta^2(k_2-k_3)~,}
\auto
$$

$$
\eqalign{K^{(4)}_1~=~-{1 \over 3}~
\sum ~ k_1^4J_2(k_1^2)k_2^2(16\pi^3)\delta^2(k_2-k_3)}
\auto
$$

$$
\eqalign{K^{(4)}_2~=~- {1 \over 2}~ \sum
\Biggl({k_1^2J_1(k_1^2)k_2^2k_3^2+
k_1^2k_3^2J_1(k_4^2)k_4^2 \over
(k_1-k_4)^2} \Biggr),}
\auto
$$

$$
\eqalign{K^{(4)}_3~=~{1 \over 2} ~\sum~
k_2^2k_4^2J_1((k_1-k_4)^2)~,}
\auto
$$
and
$$
\eqalign{K^{(4)}_4~=~{1 \over 4}~\sum~
k_1^2k_2^2k_3^2k_4^2~I(k_1,k_2,k_3,k_4), }
\auto
$$
where $J_1(k^2)$ is defined by (\ref{j1}) and
$$
\eqalign{J_2(k^2)~=~{1 \over 16\pi^3}\int d^2q {1 \over
(k-q)^2}J_1(q^2)~~~,}
\auto
$$
and
$$
\eqalign{ I(k_1,k_2,k_3,k_4)~=~{1 \over 16\pi^3}\int d^2p {1 \over
p^2(p+k_1)^2(p+k_1-k_4)^2(p+k_3)^2}.}
\auto\label{box}
$$

As for the $O(g^2)$ kernel, the only remnant of the color structure is in
the normalization factor of $N^2$. We will show in \cite{exp} that the
coefficients of the contributing $K^{(4)}_i$ and the absolute normalization
of $K^{4n}_{2,2}$ are determined directly by $t$-channel unitarity, together
with the color factors given by the group structure, apart from the
ambiguity of the magnitude of the new 1-3 coupling appearing in $K_1^{(4)}$,
$K_2^{(4)}$ and $K_3^{(4)}$. The diagrammatic representation of
${\hat{K}}^{4n}_{2,2}$ is shown in Fig.~3.1.

\vspace{.2in}

\epsffile{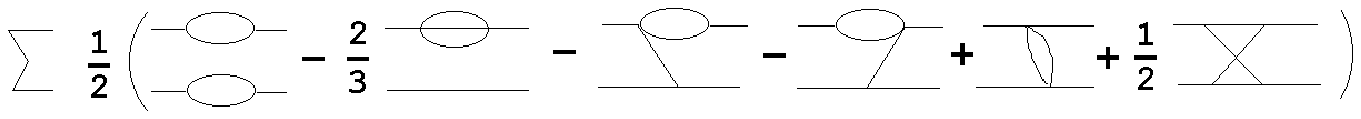}
\begin{center}
Fig.~3.1 The diagrammatic representation of ${\hat{K}}^{4n}_{2,2}$.
\end{center}

\noindent From this figure it is clear that the 1-3 vertex is the only new
ingredient of the $O(g^4)$ kernel compared to the $O(g^2)$ kernel. It's
magnitude is determined by the Ward Identity constraint that the kernel
should vanish when $k_i \to 0,~i=1,..,4$.

Diagrammatically the Ward identity is satisfied as illustrated in Fig.~3.2

\vspace{.2in}

\epsffile{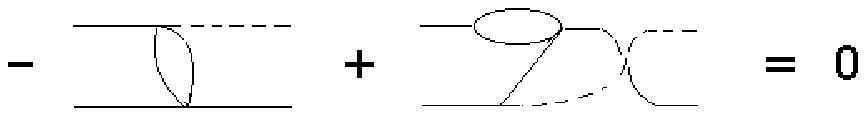}
\begin{center}
Fig.~3.2 The Ward identity constraint for $K^{4n}$.
\end{center}

\noindent and so determines the relative weight of $K_2^{(4)}$ and $K_3^{(4)}$.
Although both diagrams contain the new 1-3 vertex, $K_3^{(4)}$ contains the
square of this vertex, while $K_2^{(4)}$ contains it singly together with the
square of the 1-2 vertex. Therefore the Ward identity determines the new vertex
in terms of the square of the 1-2 vertex.

There are two infra-red finiteness requirements following from the zero
color of the kernel. These lead to three constraints, that determine the
relative weights of the remaining components. First we require that the
connected part of the kernel is infra-red finite before integration. This is
illustrated in Fig.~3.3

\medskip

\epsffile{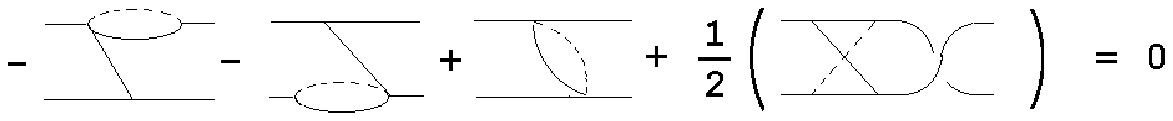}
\begin{center}
Fig.~3.3 Infra-red finiteness of the connected part of $K^{4n}$.
\end{center}

\noindent and determines $K_4^{(4)}$ relative to $K_2^{(4)}$ and $K_3^{(4)}$.
Taking the Ward identity zeroes into account, infra-red finiteness after
integration requires cancellation, by the disconnected parts, of two
divergences due to the connected part. First the poles of $K_2^{(4)}$ require
the cancellation shown in Fig.~3.4.

\medskip

\epsffile{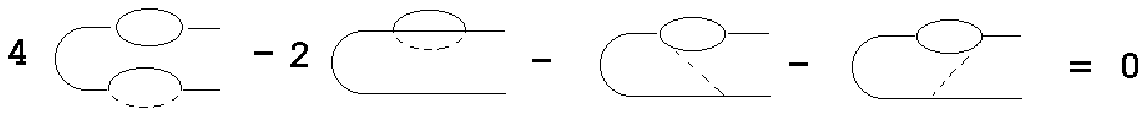}
\begin{center}
Fig.~3.4 Infra-red cancellation of the poles in $K_2^{(4)}$.
\end{center}

\noindent Secondly $K_3$ generates a divergence, when both exchanged lines
carry zero transverse momentum, which requires the cancellation shown in
Fig.~3.5.

\medskip

\epsffile{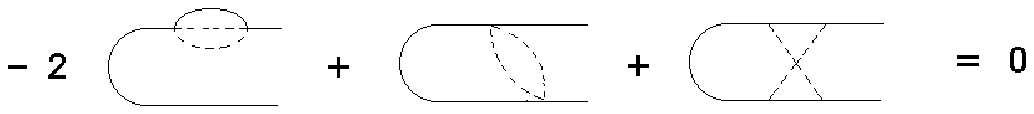}
\begin{center}
Fig.~3.5 Infra-red cancellation of divergences due to $K_3^{(4)}$.
\end{center}

\noindent This last constraint determines $K^{(4)}_1$ relative to
$K^{(4)}_2 + K^{(4)}_3 + K^{(4)}_4$ and the previous constraint then
determines the relative weight of $K^{(4)}_0$.

In our original construction of $K^{(4n)}$ in \cite{ker}, we particularly
emphasized that the relative weights of all the terms in $K^{(4n)}$ are
determined by the combination of the Ward identity constraint with the
infra-red finiteness cancellations. However, as we stated above, we show in
\cite{exp} that $t$-channel unitarity, together with the Ward identities and
color factors, also determines all coefficients. Therefore infra-red
finiteness can actually be deduced from $t$-channel unitarity, with the use
of Ward identities, as might be anticipated.

\medskip

\epsffile{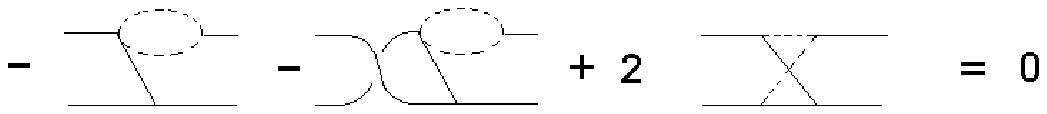}
\begin{center}
Fig.~3.6 Threshold cancellation in $K_2^{(4)}$ and $K_4^{(4)}$.
\end{center}

Finally we return to the issue of the interplay of multiplicative divergences
and explicit Ward identity zeroes. $K^{(4)}_2$ contains diagrams with this
problem. However, $K^{(4)}_4$ also has an explicit zero for each external
line which is invalidated by threshold divergences. As illustrated in
Fig.~3.6, after infra-red cancellations, the Ward identity is preserved by
cancellation of the leading threshold behaviour of $K^{(4)}_2$  and
$K^{(4)}_4$.

The most complicated part of $K^{(4n)}$ is clearly $K^{(4)}_4$ since it
contains the box diagram i.e. a loop integral with four propagators. This
diagram will actually occupy a major part of our remaining discussion. In
the next Section we describe the explicit evaluation of the integral as a
sum of logarithms.

\mainhead{4. EVALUATION OF THE BOX AS A SUM OF LOGARITHMS}

Since $I(k_1,k_2,k_3,k_4)$ is an infra-red divergent integral, to obtain an
explicit expression for it we first introduce a mass $m$ for each
propagator. We then have a standard one-loop integral which appears in many
two-dimensional field theories.

There are two basic ways of evaluating one loop n-point functions in two
dimensions. According to ref.~\cite{devega}, in $D=2$, form factors can be
reduced to a sum of 3 self energy integrals, and, in general, an n-point
function reduces to three $(n-1)$-point functions. The procedure can be
applied iteratively until one is left only with a set of self energy
integrals, each of them giving one logarithm. In the case of the box diagram
this method generates 9 logarithms ($1$ box $\to$ 3 form factors $\to$ 9
logs).

In principle one could also use the zero mode invariance of the
decomposition of ref.~\cite{devega} to reorganize the coefficients of the
logarithms and so reduce their number, but in practice this can be tedious.
Therefore, it is convenient to keep the number of logarithms as low as
possible from the beginning. As the infra-red analysis of the previous
Section shows, the box has 4 distinct infra-red divergences, each generating
a logarithm as a mass singularity. However, as we shall see, a complete
evaluation of the box involves, as a minimum, logarithms associated with all
possible kinematic thresholds. A systematic method to evaluate this minimal
number of logs has been developed long ago by K\"{a}llen and Toll \cite{KT}
and, later, in the analysis of the massive Thirring model, by Berg
\cite{Berg}.

In this last method one propagator is set on shell by the first integration and
a second is set on shell after partial fractioning. The second integration then
generates a logarithm. The remaining parts of the diagram, the ``trees'',
are factored out. The non-trivial part of the procedure is in fact the
evaluation of the tree contributions. This is rather cumbersome and requires
the introduction of dual momenta for the internal lines. In our application
we have found it convenient to re-express these dual momenta in terms of the
original momenta of the scattering diagram. We are then able to obtain
explicit, if rather lengthy, expressions. We will present most of the
details of our calculations in Appendices, but we can outline them as
follows.

In order to apply the K\"{a}llen and Toll\cite{KT} formalism directly we
evaluate integrals in the timelike region. In fact with this formalism we
obtain
\beqa
16\pi^3 J_1(k^2,m^2)&=&\int d^2 p{1\over
(p^2 + m^2)((p-k)^2 + m^2)} \nonumber \\
 &=& {i \pi\over \lambda^{1/2}(k^2,m^2,m^2)}
Log \left( {{k^2 -2 m^2 -\lambda^{1/2}(k^2,m^2,m^2)}\over {k^2 -2 m^2
+\lambda^{1/2}(k^2,m^2,m^2)}}\right). \nonumber \\
\label{j1m}
\eeqa
where now $\lambda^{1/2}(k^2,m^2,m^2)=\sqrt{k^2(k^2 - 4 m^2)}$. (To agree
directly with the analytic continuation from negative $k^2$ we should
actually evaluate the two logarithms in (\ref{j1m}) on different sheets. But
since we are only interested in the real parts of integrals as $m^2 \to 0$,
this is irrelevant for our purposes. This is discussed further in D.)

In order to discuss a planar box diagram we also, temporarily, interchange
$k_1$ and $k_4$ and take all momenta to be flowing into the diagram. We
therefore consider
$$
\eqalign{I_4 (k_1&,k_2,k_3,k_4,m^2)~=~ 16\pi^3 I (k_4,k_2,-k_3,-k_1,m^2) \cr
&=~\int d^2 p {1\over [p^2 - m^2] [(p + k_1)^2 - m^2]
[(p + k_1 + k_2)^2 - m^2][ (p - k_4)^2 - m^2]} }
\auto\label{im}
$$
We shall use the notation $p_i,~(i=1,..,4)$ for the internal momentum flowing
along the $i$-th internal line, in addition to the loop momentum $p$, and
defined such that if $p_{jk}~=~(p_j-p_k)^2$ then $p_{j,j+1}=k_j$ (with
the obvious notation that $p_5 ~\equiv p_1$). This notation is illustrated
in Fig.~4.1.

\epsffile{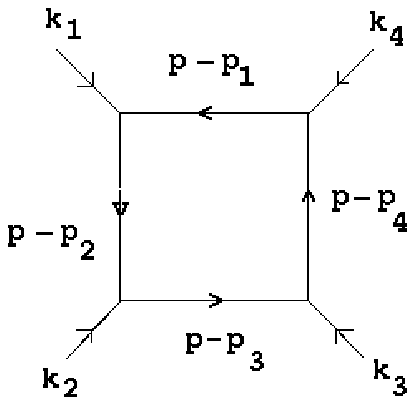}
\begin{center}
Fig.~4.1 Notation for the box diagram
\end{center}

\noindent There will be one logarithm in our result for each $p_{jk}$. In
anticipation of this we define $X_{jk}$ by
\beq
X_{j k}={p_{jk}^2 -2 m^2 -\lambda^{1/2}(p_{jk}^2,m^2,m^2)\over
p_{jk}^2 -2 m^2 + \lambda^{1/2}(p_{jk}^2,m^2,m^2)}
\label{dX}
\eeq
and write
\beq
F_{jk}~\equiv~F(p_{jk}^2,m^2)~=~
{i\pi\over \lambda^{1/2}(p_{jk}^2,m^2,m^2)}Log X_{jk}.
\label{dF}
\eeq

We similarly identify trees $A_{jk}$ by the indices of the lines which have
been set on shell. This is illustrated in Fig.~4.2.

\vspace{.2in}

\epsffile{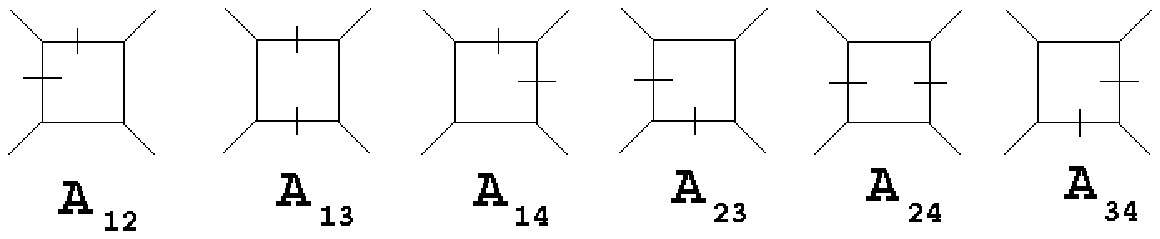}
\begin{center}
Fig.~4.2 Tree diagrams obtained by putting on-shell the crossed lines.
\end{center}

\noindent To evaluate the trees, we need an explicit expression for the loop
momentum $p$ which satisfies
\beqa
&& (p-p_j)^2 -m^2=0 \nonumber \\
&& (p-p_k)^2 - m^2=0.
\eeqa
These two conditions are solved in \cite{KT} by the vectors
\beq
p=q_{jk}^{\pm}={1\over 2}
\left( p_j + p_k \pm p^d_{kj}{\lambda^{1/2}(p_{jk}^2,m^2,m^2)\over p_{jk}^2}
\right).
\label{dua}
\eeq
where we have introduced dual vectors $p^d$ satisfying the conditions
\beq
{p^d}_{jk}\cdot p_{jk}=0, \,\,\,\,\,\,\,{p^d}_{jk}^2=-{p_{jk}}^2
\eeq

Since $p_{j,j+1}=k_j$, we can go one step further compared to \cite{KT}
by noticing that the dual momenta can be expressed in terms of momenta that
are simply (dual) orthogonal to the external lines. In particular we can take

\beqa
&& {k^d}_1={\epsilon(n_{12})\over \sqrt{(k_1\cdot k_2)^2 - k_1^2 k_2^2}}
\left( k_1 k_1\cdot k_2 - k_2 k_1^2\right) \nonumber \\
&& {k^d}_2={\epsilon(n_{21})\over \sqrt{(k_1\cdot k_2)^2 - k_1^2 k_2^2}}
\left( k_2 k_1\cdot k_2 - k_1 k_2^2\right) \nonumber \\
&& {k^d}_3={\epsilon(n_{34})\over \sqrt{(k_3\cdot k_4)^2 - k_3^2 k_4^2}}
\left( k_3 k_3\cdot k_4 - k_4 k_3^2\right) \nonumber \\
&& {k^d}_4={\epsilon(n_{43})\over \sqrt{(k_3\cdot k_4)^2 - k_3^2 k_4^2}}
\left( k_3 k_3\cdot k_4 - k_4 k_3^2\right), \nonumber \\
\label{duals}
\eeqa
where we have defined
\beqa
n_{12}=k_1^1 k_2^0 - k_1^0 k_2^1 ~,~~~~~
n_{34}=k_3^1 k_4^0 - k_3^0 k_4^1 ~,
\eeqa
and $n_{21}=-n_{12}$, $n_{34}=-n_{43}$. $\epsilon(x)=\theta(x)-\theta(-x)$
is the sign function. We have the relations
\beqa
&&  n_{12}~=~\sqrt{(k_1\cdot k_2)^2 - k_1^2 k_2^2}~
=~\sqrt{\lambda(s,k_1^2,k_2^2)}
\nonumber \\
&&n_{12}~=~\sqrt{(k_3\cdot k_4)^2 - k_3^2
k_4^2}~=~\sqrt{\lambda(s,k_3^2,k_4^2)}
\nonumber \\
\eeqa

We now define
\beq
A_{jk}={1\over 2}\left( {A^+}_{jk} + {A^-}_{jk}\right).
\eeq
where
\beq
{1\over A_{jk}^{\pm}}=
\prod_{l\neq j,k}\left( (q_{jk}^\pm - p_l)^2 + m^2
\right)
\label{prod}
\eeq
In a general n-point function the product in (\ref{prod}) would run over the
$n-2$ propagators which are left off-shell. Then the full expression for an
n-point diagram (here denoted by $I_n$) is given by
\beq
I_n=\sum_{j\,\,<\,\,k}A_{jk} F_{jk}.
\label{In}
\eeq
including, of course, $n=4$. The number of coefficients generated is
$n(n-1)/2$ for an $n$-point function.

In the specific case of the planar box diagram this method gives (to
simplify the evaluation for the next Section we take $k_3$ and $k_4$ to be
flowing out)
$$
\eqalign{
I_4(k_1,k_2&,k_3,k_4)~~\biggl(~\equiv ~I(-k_4,k_2,k_3,-k_1)~\biggr) \cr
=~~& A_{12}F\left(k_3^2,m^2\right)+
 A_{13}F\left( (k_1 + k_2)^2,m^2)\right)
+ A_{14}F\left(k_4^2,m^2\right) \cr
+& A_{23}F\left(k_2^2,m^2\right)
+A_{24}F\left((k_2-k_3)^2,m^2\right) + A_{34}F\left(k_1^2,m^2\right) }
\auto\label{I4}
$$
where F is given by (\ref{dF}).

Further discussion of the the results with $m^2$ kept finite can be found in
the Appendix. Clearly in the limit $m^2 \to 0$ only the $X_{jk}$ are
divergent. The trees, i.e. the $A_{jk}$ simplify considerably and we give
examples of their complete expression here. Note that although the
$A_{j,k}^{\pm}$ are initially defined separately as a product of simple
denominators, combining them leads to relatively complicated expressions.
For convenience we define
\beq
A_{ij}={a_{ij}\over b_{ij}}
\eeq
Examples of the $a_{ij}$ and $b_{ij}$ are
\beqa
a_{12} = &&\biggl[ {{{ k_1} \cdot { k_2}}^2} - { k_1}^{2}\,{ k_2}^{2}
\biggr]\nonumber \\
&&\,\,\,\,\,\,\,\,\,\,\,\times\biggl[{{{ k_1} \cdot { k_2}}^2} -
   { k_1} \cdot { k_2}\,{ k_1} \cdot { k_3} -
   { k_1}^{2}\,{ k_2}^{2} + { k_1}^{2}\,{ k_2} \cdot { k_3}\biggr.
\nonumber \\
&&\biggl. +  ( { k_1} \cdot { k_2} + { k_2}^{2} ) \,
    ( { k_1} \cdot { k_2} - { k_1} \cdot { k_3} + { k_2}^{2} -
      2\,{ k_2} \cdot { k_3} + { k_3}^{2} )\biggr]
\eeqa
\beqa
b_{12} = &&\biggl[ -{{{ k_1} \cdot { k_2}}^2} + { k_1}^{2}\,{ k_2}^{2} +
   {{( { k_1} \cdot { k_2} + { k_2}^{2} ) }^2}\biggr]
\nonumber \\
&&\times\left.\biggl[ -{{( {{{ k_1} \cdot { k_2}}^2} -
         { k_1} \cdot { k_2}\,{ k_1} \cdot { k_3} -
         { k_1}^{2}\,{ k_2}^{2} +
         { k_1}^{2}\,{ k_2} \cdot { k_3} ) }^2}\right.\nonumber \\
&&\left. +
   ( {{{ k_1} \cdot { k_2}}^2} - { k_1}^{2}\,{ k_2}^{2})
      ) \,{{( { k_1} \cdot { k_2} - { k_1} \cdot { k_3} +
         { k_2}^{2} - 2\,{ k_2} \cdot { k_3} + { k_3}^{2} ) }^2}
\right.\biggr]\nonumber \\
\eeqa
\beqa
a_{23} = &&\biggl[ {{{  k_1} \cdot {  k_2}}^2} - {  k_1}^{2}\,{
k_2}^{2}\biggr]
\nonumber\\
&&\,\,\,\,\,\,\,\,\times\biggl[ -( {  k_1} \cdot {  k_3}\,{  k_2}^{2} )  +
   {  k_1} \cdot {  k_2}\,{  k_2} \cdot {  k_3} +
   ( {  k_1}^{2} + {  k_1} \cdot {  k_2} ) \,
    ( -{  k_2} \cdot {  k_3} + {  k_3}^{2} ) \biggr]\nonumber\\
\eeqa
\beqa
b_{23} = &&\biggl[ -{{{  k_1} \cdot {  k_2}}^2} +
   {{( {  k_1}^{2} + {  k_1} \cdot {  k_2} ) }^2} +
   {  k_1}^{2}\,{  k_2}^{2}\biggr]\nonumber\\
&&\times\biggl[ -{{[ -( {  k_1} \cdot {  k_3}\,{  k_2}^{2} )  +
         {  k_1} \cdot {  k_2}\,{  k_2} \cdot {  k_3} ] }^2} +
   ( {{{  k_1} \cdot {  k_2}}^2} - {  k_1}^{2}\,{  k_2}^{2} )
\,{{( -{  k_2} \cdot {  k_3} + {  k_3}^{2} ) }^2}\biggr]\nonumber \\
\eeqa
\beqa
a_{34} = &&\biggl[ {{{  k_3} \cdot {  k_4}}^2} - {  k_3}^{2}\,{  k_4}^{2}
 \biggr]~\biggl[\left( [ - {  k_2} \cdot {  k_4}\,{  k_3}^{2}   +
      {  k_2} \cdot {  k_3}\,{  k_3} \cdot {  k_4}]  \,
    [ -( {  k_1} \cdot {  k_4} +
           {  k_2} \cdot {  k_4} ) \,{  k_3}^{2}\right. \nonumber \\
&& \left.+ ( {  k_1} \cdot {  k_3} + {  k_2} \cdot {  k_3} ) \,
       {  k_3} \cdot {  k_4} ]\right) +\left(
   ( {  k_2}^{2} - {  k_2} \cdot {  k_3} ) \,
    ( {  k_1}^{2} + 2\,{  k_1} \cdot {  k_2} -
      {  k_1} \cdot {  k_3} \right.\nonumber\\
&& \left. + {  k_2}^{2} - {  k_2} \cdot {  k_3})
      \,( {{{  k_3} \cdot {  k_4}}^2} -
      {  k_3}^{2}\,{  k_4}^{2} )\right) \biggr]\nonumber \\
\eeqa
The remaining $a_{ij}$ and $b_{ij}$ are given in Appendix B.

Since $K^{(4)}_2$ and $K^{(4)}_3$ involve only $J_1$ it is clear that
by utilising (\ref{j1m}), (\ref{dX}), (\ref{dF}) and (\ref{I4}) and taking
the $m^2 \to 0$ limit in all logarithms, we can obtain an analytic
expression for the the full connected part of the kernel. We will not write
it out in full. Instead we proceed directly to the forward kernel which is
very much simpler. We also will not consider the full disconnected part here.
We
will discuss this explicitly for the case of the forward kernel in the next
Sections.

\mainhead{5. THE $O(g^4)$ PARTON KERNEL}

As we have discussed in Section 2, it is the forward kernel
$K^{(4n)}_{2,2}(q-k,k,k',q-k')$ which appears in the parton evolution equation.
To evaluate the contribution of $K^{(4)}_4$ we require
$I(k,k')=I_4(k',k,k',k)$.
The simplification of the $X_{jk}$ as $m^2 \to 0$ gives
$$
\eqalign{ 4{\pi}^2& I[k,k']=~{A_{12} \over {k'}^2} Log[{ k'}^2/m^2] +
{A_{23} \over k^2} Log[k^2/m^2] + {A_{34} \over {k'}^2} Log[{k'}^2/m^2] \cr
&+ {A_{13} \over (k+k')^2}  Log[(k+k')^2/m^2] +
{A_{14}  \over k^2} Log[k^2/m^2] + {A_{24} \over (k-k')^2}
Log[(k-k')^2/m^2]}
\auto\label{If}
$$
The complicated expressions for the $A_{jk}$ given in the previous Section
simplify enormously when $q=0$ and we obtain
$$
\eqalign{  &A_{12} = {k^2 - {k'}^2 \over k^2 (k+k')^2(k-k')^2}
{}~~~~~~~~~~~~~~~~~~~~~~A_{13}={1\over k^2 {k'}^2}\cr
&A_{14}={{k'}^2-k^2\over k'^2 (k+ k')^2 (k-k')^2}
{}~~~~~~~~~~~~~~~~~~~~~A_{23}={k'^2-k^2\over k'^2 (k+k')^2 (k-k')^2}\cr
&A_{24}={1\over k^2 {k'}^2}
{}~~~~~~~~~~~~~~~~~~~~~~~~~~~~~~~~~~~~~~~~A_{34}=
{k^2- {k'}^2 \over k^2 (k+k')^2(k-k')^2}}
\auto
\label{f1}
$$
Note that the $m^2$ divergences multiplying $A_{23}$ and $A_{14}$, and
$A_{34}$ and $A_{12}$ cancel pairwise. The remaining divergences have to
cancel with the remaining terms of the kernel.

{}From (\ref{If}) and (\ref{f1}) we have that, as $m^2 \to 0$,
$$
\eqalign{K^{(4)}_4 ~\to ~ {-~k^2 {k'}^2 \over 8\pi^2}&\Biggl(
{ 2({k'}^2 -k^2) \over (k+k')^2 (k-k')^2}
Log\Biggl[{{k'}^2\over k^2} \Biggr] \cr
&+ {1\over (k-k')^2} Log\Biggl[{ (k-k')^2\over m^2}\Biggr]
{}~+~{1\over (k+k')^2} Log\Biggl[{ (k +k')^2\over m^2}\Biggr]
\Biggr)~ .}
\auto\label{bm2}
$$
$K^{(4)}_3$ simply gives a contribution of the same form as the last
two terms in (\ref{bm2}), i.e. as $m^2 \to 0$
\beqa
\tilde{K}_3~\to~{k^2 {k'}^2 \over 8\pi^2} \left(
{1\over (k-k')^2} Log\left[{ (k-k')^2\over m^2}\right]
+ { 1\over (k+k')^2} Log\left[{ (k+k')^2\over m^2}\right]\right)
\label{k_30}
\eeqa
Similarly $K^{(4)}_2$ gives
$$
\eqalign{
\tilde{K_2} \to {-k^2 {k'}^2 \over 8\pi^2 } & \Biggl(
{1 \over (k-k')^2} (Log\Biggl[{ k^2\over m^2}\Biggr] +
Log\Biggl[{ {k'}^2\over m^2}\Biggr] ) \cr
 + & { 1\over (k+k')^2} (Log\Biggl[ { k^2\over m^2} \Biggr] +
Log\Biggl[{ {k'}^2 \over m^2} \Biggr] ) \Biggr).}
\auto\label{k_20}
$$

The infra-red finiteness of the connected part $K^{(4n)}_c ~=~ K^{(4)}_2 +
K^{(4)}_3 + K^{(4)}_4$ is now apparent and we can write
$$
\eqalign{K^{(4n)}_c~&=~{1 \over 8\pi^2}
\Biggl( {k^2{k'}^2 \over (k-k')^2}Log\left[{(k-k')^4 \over k^2{k'}^2 }\right]
{}~+~ {k^2{k'}^2 \over (k+k')^2} Log\left[{(k+k')^4 \over k^2{k'}^2 }\right]
\Biggr)\cr
&~~~~~-~~~~
\Biggl( {2 k^2{k'}^2 (k^2 - {k'}^2) \over (k-k')^2(k+k')^2}Log\left[{k^2 \over
{k'}^2}\right] \Biggr) \cr
&=~~~~ \Biggl( ~~{\cal K}_1~~\Biggr) ~~-~~\Biggl( ~~{\cal K}_2~~\Biggr)
. }
\auto\label{4nc}
$$
Only ${\cal K}_1$ gives infra-red divergences (at $k'= \pm k$) when
integrated over $k'$. The infra-red analysis of Section 3 showed that these
divergences are cancelled by $K_0$ and $K_1$. (Note that the normalization
of ${\cal K}_1$ and ${\cal K}_2$ differs from that in \cite{spec}.)

We emphasized in \cite{spec} that ${\cal K}_2$ has a number of attractive
properties which make it interesting to study separately and we shall see
this in the following. However, we should also note the following
inter-relation between the two components. If we consider the limit $k^2 \to
0$ we find
$$
{\cal K}_1 ~\to~ {k^2 \over 4\pi^2} Log\left[{{k'}^2 \over k^2 }\right]
{}~~,~~~~~~~~~~~~~~~
{\cal K}_2 ~\to~ - {k^2 \over 4\pi^2} Log\left[{{k'}^2 \over k^2 }\right]~~,
\auto
$$
and so there is a cancellation between ${\cal K}_1 $ and ${\cal K}_2 $. This
cancellation is a consequence of the Ward identity cancellation between
$K^{(4)}_2$ and $K^{(4)}_4$ discussed in Section 3 and illustrated by
Fig.~3.6. Consequently the Ward identity property relates ${\cal K}_1 $ and
${\cal K}_2 $ even though their infra-red properties allow them to be
separated.

Apart from the logarithmic factors, ${\cal K}_1$ has the same structure as
the forward (connected) $O(g^2)$ kernel. Indeed, we now show that it is
directly related to the square of the $O(g^2)$ kernel, evaluated in the
forward direction. We show that (in the forward direction)
$$
\eqalign{ \hat{{\cal K}}_0~+~ \hat{{\cal K}}_1 ~=~
{1 \over 4} \biggl(\hat{K}^{(2)}_{2,2}\biggr)^2~, }
\auto\label{k_2s}
$$
where ${\cal K}_0$ represents the sum of the disconnected parts $K^{(4)}_0$
and $K^{(4)}_1$. (\ref{k_2s}) will provide a simple determination of the
eigenvalue spectrum of ${\cal K}_0~+~{\cal K}_1$.

We derive (\ref{k_2s}) diagrammatically. The full square of
$\hat{K}^{(2)}_{2,2}$ is given in Fig.~5.1.

\vspace{.2in}

\epsffile{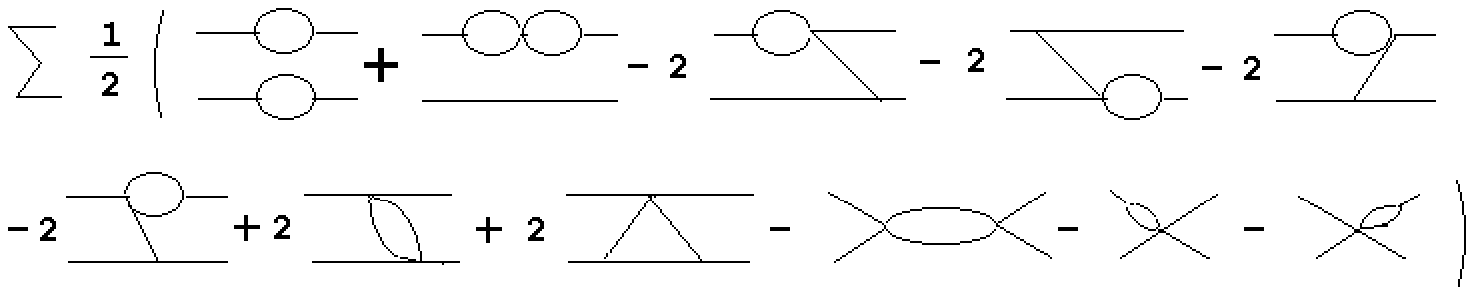}
\begin{center}
Fig.~5.1 The full square of $\hat{K}^{(2)}_{2,2}$
\end{center}

\noindent In the forward direction various distinct diagrams become equal.
This is shown in Fig.~5.2.

\vspace{.2in}

\epsffile{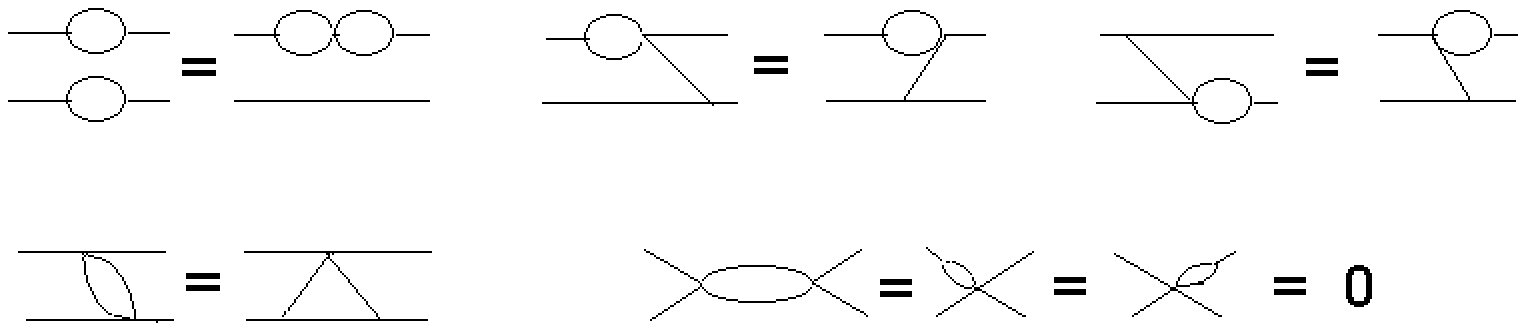}
\begin{center}
Fig.~5.2 Equalities in the forward direction.
\end{center}

\noindent and so $(\hat{K}^{(2)}_{2,2})^2$ simplifies to the result shown in
Fig.~5.3.

\vspace{.2in}
\epsffile{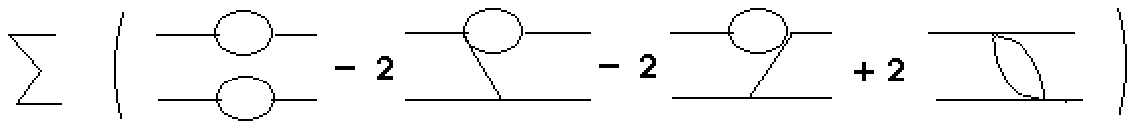}
\begin{center}
Fig.~5.3 Representation of $(\hat{K}^{(2)}_{2,2})^2$ in the forward direction.
\end{center}

\noindent From this we see that if we utilise (\ref{k_30}) and (\ref{k_20}),
then (\ref{k_2s}) holds apart from the contribution of the disconnected parts.
However, in Appendix D we show, using the dimensional regularization which
we discuss just below, that
$$
4k^2[J_2(k^2)]~=~3 [k^2J_1(k^2)]^2
$$
and this is sufficient to show that (\ref{k_2s}) is also valid when the
disconnected parts are included.

To introduce a complete dimensionally regularized form for $K^{(4n)}(k,k')$
we replace each $Log[k^2]$ using the dimensionally regularized form of
$J_1(k^2)$, i.e.
\beqa
Log[k^2]~=~8\pi^2J_1(k^2)~=~
{k^2 \over 2\pi}\int {d^D q\over q^2 (k-q)^2}~=~\eta[k^2]^{D/2~-1}
\eeqa
where
$$
\eqalign{\eta={\Gamma[2-D/2]\Gamma[D/2-1]^2\over 2 \Gamma[D-2]}~
\centerunder{$\longrightarrow$} {\raisebox{-5mm} {$D \to 2$}}~{2 \over
(D-2)} }
\auto
$$
Notice that this is possible (in $D=2 +\epsilon$ dimensions)
$only$ because we have shown the cancellation of the mass singularities in
eq.~\ref{4nc}. In fact, in the limit of $\epsilon\to 0$, a suitable choice
of the dimensional regularization scale $\mu$ allows us to reproduce the
same expressions for the components of the kernel regulated by a cutoff $m$.

We can then write the complete normalised kernel in D dimensions as
$$
\eqalign{{1 \over (g^2N)^2} K^{(4n)} ~=~ \Biggl( ~{\cal K}_0~\Biggr)
{}~+~\Biggl( ~{\cal K}_1~\Biggr) ~-~\Biggl( ~{\cal K}_2~\Biggr)
, }
\auto\label{4n}
$$
where now
$$
\eqalign{ {\cal K}_0~=~{1 \over 8\pi}{\eta}^2(k^2)^D \biggl(\delta^2(k-k')
+\delta^2(k-k')\biggr)}
\auto
$$
$$
\eqalign{ {\cal K}_1~=~&{\eta \over 8\pi^2} \biggl( 2 k^2
{k'}^2([(k'-k)^2]^{D/2-2}~+~[(k+k')^2]^{D/2-2})~\cr
&-(k^2 [{k'}^2]^{D/2} ~+~[k^2]^{D/2} {k'}^2)\bigl({1 \over (k'-k)^2}
{}~+~ {1 \over(k'+k)^2}\bigr)\biggr)}
\auto
$$
and
\beqa
{\cal K}_2~=~{\eta \over 4\pi^2} ~ {k^2 {k'}^2 (k^2-{k'}^2)\over
(k+k')^2 (k-k')^2}\left( (k^2)^{D/2~-1} -
({k'}^2)^{D/2~-1}\right).
\eeqa

\mainhead{6. IMPACT PARAMETER REPRESENTATION}

In this Section we give impact parameter repesentations, for components of the
forward kernel, that we anticipate will be useful in further studies. In
particular with respect to conformal symmetry and holomorphic factorization
properties. We limit our discussion to the dimensionally regulated form of
the kernel (in $D=2 +\epsilon$ dimensions) and do not specialize our results
to $D=2$.

The expression for ${\cal K}_1(x,y) $ turns out to be much simpler than the
corresponding expression for ${\cal K}_2 (x,y)$, which can be written down
 in terms of a Gegenbauer expansion. For this purpose
we have used a generalized
expression for a 3-point function given in ref.~\cite{davi}, and which is
briefly described in Appendix F. The Gegenbauer expansion is a particular
case of the general result given in \cite{davi}, and can be derived quite
simply by expanding euclidean propagators (in $D$-dimensions) in the base
of Gegenbauer polynomials \cite{chet}.

We define
\beqa
&& {\cal K}_1(x,y)\equiv
{1\over (2 \pi)^{2 D}}\int d^D\,k\,d^D\,k' \,e^{i k\cdot x} e^{i k\cdot y}
{\cal K}_1(k,k').\nonumber
\eeqa
It is convenient to also define
\beqa
g_1(k,k')= {2 k^2 k'^2\over [(k'-k)^2]^{2-D/2}} ~, ~~~~~~~
g_2(k,k')={k^2 [k'^2]^{D/2}\over (k'-k)^2} ~,
\eeqa
and let $g_1(x,y)$ and $g_2(x,y)$ denote their corresponding expressions
in impact parameter space. Noting that $g_1(x,y)$ is defined only in a
distributional sense, we can write
\beqa
&&{\cal K}_1(x,y)={\eta\over 8 \pi^2}\left( g_1(x,y) +g_1(x,-y)\right.
 \nonumber \\
&&\left. -(g_2(x,y) +g_2(y,x)) -(g_2(-x,y) +g_2(-y,x))\right)
\eeqa
and show that to $O(1/\epsilon)$
\beq
{\eta\over 8 \pi^2}(g_1(x,y) +g_1(x,-y)) + {\cal K}_0(x,y)=0
\eeq
As an explicit expression for $g_2(x,y)$ we can write
\beqa
&& g_2(x,y)={-2^{D-2}\over \pi^D}{\Gamma[D]\Gamma[D/2-1]\over \Gamma[-D/2]}
\Box_x {1\over [x_+^2]^D [x^2]^{D/2-1}}.\nonumber \\
\eeqa
where $\Box_x\equiv {\partial\over \partial_\mu}{\partial\over \partial^\mu}$ .
and
\beq
x_+=x+y,\,\,\,\,\,\,\,\,\,\, x_-=x-y.
\eeq
For ${\cal K}_2$ we can obtain a more complete expression. We write
\beqa
{\cal K}_2(x,y)={\eta\over 4 \pi^2}\left( f_1(x,y)- f_2(x,y)
  + f_1(y,x) -f_2(y,x) \right)
\eeqa
with $f_1(x,y)$ and $f_2(x,y)$ being, respectively, the impact parameter
representations of
\beqa
f_1(k,k')={k'^2[k^2]^{D/2+1}\over (k+k')^2 (k-k')^2}~,~~~~~
f_2(k,k')={[k'^2]^2 [k^2]^{D/2}\over (k+k')^2 (k-k')^2}~.
\eeqa
We obtain
\beqa
 f_1(x,y)= -{1\over \pi^{3 D/2}}{\Gamma[D/2-1]^2\Gamma[D+1]\over
\Gamma[-D/2-1]} 2^{3 D -6}\Box_y Y(D+1,D/2-1,D/2-1|x_+,x_-) \nonumber \\
\eeqa
and
\beqa
f_2(x,y)={1\over \pi^{3 D/2}}{\Gamma[D/2-1]^2\Gamma[D]\over \Gamma[-D/2]}
2^{3 D-8} (\Box_y)^2 Y(D,D/2-1,D/2-1|x_+,x_-)
\eeqa
where
\beq
Y(\sigma_1,\sigma_2,\sigma_3|x_+,x_-)=\int
{d^D~z_1\over [z_1^2]^{\sigma_1}[(x_+ + z_1)^2]^{\sigma_2}
[(x_- + z_1)^2]^{\sigma_3}}
\label{vertex}
\eeq
is a generalized vertex function in impact parameter space.

In ref.~\cite{davi} it is shown that eq.~(\ref{vertex}) can be re-expressed
in terms of a Mellin-Barnes transform
\beqa
&&Y(\sigma_1,\sigma_2,\sigma_3,x_+,x_-)={\pi^{D/2}i^{1-D}
[x_+^2]^{D/2-\sigma_1-\sigma_2-\sigma_3}\over
(2 \pi i)^2\Gamma[\sigma_1]\Gamma[\sigma_2]
\Gamma[\sigma_3]\Gamma[D-\sigma_1-\sigma_2-\sigma_3] 4 \pi^2}\nonumber \\
&&\times \int_{-i\infty}^{i\infty}~\int_{-i\infty}^{i\infty}dsdt
 {\cal X}^s {\cal Y}^t\Gamma[-s]\Gamma[-t]\nonumber \\
&&\times\Gamma[D/2-\sigma_1-\sigma_3-t]\Gamma[\sigma_3 +s +t]
\Gamma[\sigma_1 +\sigma_2 + \sigma_3 - D/2 + s +t),\nonumber \\
\label{mell}
\eeqa
with
\beq
{\cal X}={4 y^2\over x_+^2},\,\,\,\,\,\, {\cal Y}={x_-^2\over x_+^2},
\eeq
and the final result expressed in terms of generalized hypergeometric
functions (Appel functions) of two variables ${\cal X},\,\,{\cal Y}$
(see ref.~(\cite{davi})
\beq
F_4\left[a,b,c,d|{\cal X},{\cal Y}\right] =
\sum_{j=0}^{\infty}\sum_{l=0}^{\infty}{{\cal X}^j {\cal Y}^l
(a)_{j+l} (b)_{j+l}\over
j!\,\, l!\,\,\, (c)_{j} (d)_l}
\eeq
and
\beq
(a)_j={\Gamma[a+j]\over \Gamma[j]}
\eeq
denotes the Pochhammer symbol.

In our case, however, we can proceed in an alternative way.
In the particular case $\sigma_2=\sigma_3=D/2-1\equiv \lambda$, $Y$ can be
rewritten as a Gegenbauer expansion. For this purpose we introduce the
Gegenbauer polynomials
\beqa
 C_n^\lambda(x)&=&{(-2)^{-n}(1-x^2)^{-\lambda + 1/2}\Gamma[n + 2 \lambda]
\Gamma[\lambda + 1/2]\over n! \Gamma[2 \lambda]\Gamma[\lambda +1/2+n]}
{d^n\over d x^n}(1-x^2)^{n +\lambda -1/2} \nonumber \\
&=&\sum_{k=0}^{[n/2]}{(-1)^k\Gamma[\lambda + n -k](2 x)^{n-2 k}\over
k! (n-2 k)! \Gamma[\lambda]}
\eeqa
($[n/2]$ denotes the integer part of $n/2$) and by using the formulas in
Appendix F we obtain
\beq
Y(\sigma_1,D/2-1,D/2-1|x_+,x_-)={ 2 \pi^{\lambda + 1}\over \Gamma[\lambda +1]}
{\lambda\over \lambda +n}\sum_n C_n^\lambda(\hat{x}_+\cdot\hat{x}_-)
{\cal J}_n(x_+,x_-), \nonumber \\
\eeq
where we have defined $\hat{x}\equiv x^{\mu}/x^2$. An expression for
${\cal J}_n(x_+,x_-)$ can be found in the Appendix.

\mainhead{7.  SPECTRUM OF THE PARTON KERNEL}

In this Section we present results regarding the spectrum of the forward
kernel.
We use the complete set of orthogonal eigenfunctions, $\phi_{\mu,n}$, defined
in (\ref{eif}). From (\ref{k_2s}), it follows that the eigenvalue spectrum of
${\cal K}_0 + {\cal K}_1 $ is simply
$$
{1 \over \pi} [\chi(\nu,n)]^2
\auto
$$
where, again $\mu=1/2~+i\nu$, and $\chi(\nu,n)$ is given by (\ref{ei2}). If
we write $\Lambda(\nu,n)$ for the eigenvalues of ${\cal K}_2$ then the
complete spectrum of $\hat{K}^{(4n)}$ is given by $ N^2g^4 {\cal E}(\nu,n)$
with
$$
{\cal E}(\nu,n)~=~{1 \over \pi} [\chi(\nu,n)]^2~-~\Lambda(\nu,n) ~.
\auto
$$
The reggeon Green function solution of (\ref{ome}) with $K^{(4n)}$ added to
the BFKL kernel $K^{(2)}$ is, when $q=0$, a simple modification of (\ref{gre})
i.e.
$$
F(k,k')~=~\sum_{n=-\infty}^{\infty}{e^{in(\theta-\theta')} \over 4\pi^2}
\int { d\nu~~\bigl(k^2/{k'}^2\bigr)^{i\nu} \over \omega ~-~{Ng^2 \over
2\pi^2}~\chi(\nu,n)~-~{N^2g^4 \over 16\pi^3}~{\cal E}(\nu,n)}
\auto\label{gre1}
$$

To evaluate the spectrum of ${\cal K}_2$ we write
\beqa
 {\cal K}_2 \otimes \phi_{\mu,n} &=& {\cal K}_2^1 \otimes \phi_{\mu,n} -
{\cal K}_2^2 \otimes \phi_{\mu,n} \nonumber \\
&=& \lambda_1(\mu,n) \phi_{\mu,n} -
\lambda_2(\mu,n) \phi_{\mu,n} \nonumber \\
&=& \lambda(\mu,n) \phi_{\mu,n},
\eeqa
where
\beq
{\cal K}_2^1 \otimes \phi_{\mu,n}
=~{\eta \over 4 \pi^2}
\int {d^D k' \over ({k'}^2)^2} {(k^2)^{D/2} {k'}^2 (k^2-{k'}^2)
\phi_{\mu,n}(k')\over (k-k')^2(k+k')^2 ,}
\label{e1}
\eeq
and
\beq
{\cal K}_2^2 \otimes \phi_{\mu,n}
=~{\eta \over 4\pi^2}
\int {d^D k' \over ({k'}^2)^2} { k^2 ({k'}^2)^{D/2} (k^2-{k'}^2)
\phi_{\mu,n}(k')\over (k-k')^2(k+k')^2 .}
\label{e2}
\eeq
We embed the eigenfunctions in a D-dimensional angular space parameterized
by $(\theta_1,\theta_2,...,\theta_{D-1})$ by assuming that $\theta\equiv
\theta_{D-1}$. The only non trivial angular integral is then the following
one (for simplicity we write $\theta_{D-1}\equiv\theta$)
\beqa
&&I_{\chi}[n]\equiv \int_{0}^{2 \pi} d\theta{ e^{i n \theta}
\over 1- z(k,k')sin^2\,\,(\theta- \chi)}\nonumber \\
&&z[{k,k'}]=-{4 k^2{k'}^2\over (k^2 -{k'}^2)^2}
\eeqa
where $cos\chi= k\cdot \hat{x}$ and $cos\theta ={k'}\cdot \hat{x}$.
We turn the angular integral into a complex line integral on the
circle and we get by residui (for $n>\,\,-1$)
\beqa
&& I_{\chi}[n]=-4 i e^{i n \chi} \oint d w {w^{n+1}\over z w^4
+2 (2 - z)w^2 +z} \nonumber \\
&&=\,\,\,\,\,\,2 \pi\delta_{n,2 M} e^{i n \chi}
\left({k^2-{k'}^2\over k^2 + {k'}^2}\right)
\left[ \left({k\over k'}\right)^n \Theta[k'-k] -
\left( {k'\over k}\right)^n\Theta[k-k']\right]. \nonumber \\
\label{Ith}
\eeqa
$2 M$ here is an even integer - this will be important in the following. For
$n< -1$ the integral has additional poles, however it is easy to show that
$I[-n]=I[n]$ for any integer $n$. The remaining radial integral (in ${k'}^2$)
is therefore symmetric for $n\to -n$.

$I_{\chi}[n]$ is symmetric under the exchange of $k$ and $k'$, and also is
invariant under $k \to 1/k,~ k' \to 1/k'$. This last invariance is sufficient
to show from (\ref{e1}) and (\ref{e2}) that
$$
\eqalign{ \lambda(\mu~,n) ~=~\lambda(1 -\mu~,n)}
\auto\label{sym}
$$
Using (\ref{Ith}) we obtain from (\ref{e1}) and (\ref{e2}) that, as $D \to
2$,
\beqa
\lambda_1(\mu,n)~\to~{\eta \over 4\pi^2}
{\pi^{D/2}\over \Gamma[D/2]} \biggl(\beta\bigl(|n|/2 +D/2 +\mu - 1\bigr)
{}~-~\beta\bigl(|n|/2-D/2-\mu + 2\bigr) \biggr),
\eeqa
and
\beqa
\lambda_2(\mu,n)~\to~{\eta \over 4\pi^2}
{\pi^{D/2}\over \Gamma[D/2]} \biggl(\beta\bigl(|n|/2 +D +\mu - 2\bigr)
{}~-~ \beta\bigl(|n|/2-D -\mu + 3\bigr) \biggr),
\eeqa
where $\beta(x)$ is the incomplete beta function, i.e.
$$
\eqalign{
\beta(x)~&=~\int^1_0 dy~y^{x -1}[1+y]^{-1} \cr
&=~{1\over 2}\biggl(\psi\bigl({x+1\over 2}\bigr) -
\psi\bigl({x\over 2}\bigr)\biggr), }
\auto
$$
$\lambda_1(\mu,n)$ and $\lambda_2(\mu,n)$ are separately singular at $D=2$,
but $\lambda(\mu,n)$ is finite, and writing $\Lambda(\nu,n)~\equiv~
\lambda( {1 \over 2} + i\nu~,n)$, we obtain
$$
\eqalign{ \Lambda(\nu,n) ~=~-~{1 \over 4\pi}
\biggl(\beta'\bigl({|n| + 1\over 2} +
i\nu\bigr)
{}~+~\beta'\bigl({|n| + 1 \over 2} -i\nu\bigr)\biggr). }
\auto\label{lam}
$$
Since
$$
\eqalign{\beta'(x)~=~{1\over 4}\biggl(\psi'\bigl({x+1\over 2}\bigr) -
\psi'\bigl({x\over 2}\bigr)\biggr) , }
\auto\label{ps1}
$$
and
$$
\eqalign { \psi'(z)~=~\sum_{r=0}^{\infty} {1
\over (r+z)^2}, }
\auto\label{ps2}
$$
it follows that $\beta'(x)$ is a real analytic function and so, from
(\ref{lam}), the eigenvalues $\Lambda(\nu,n)$ are all real.

We can also show from (\ref{lam}) that the eigenvalues $\Lambda(\nu,n)$ have
the same holomorphic factorization property that we discussed for the leading
order eigenvalues $\chi(\nu,n)$ in Section 2. That is we can write
$$
\Lambda(\nu,n)~=~{\cal G}\bigl[m(1-m)\bigr]~+
{}~{\cal G}\bigl[\tilde{m}(1-\tilde{m})\bigr]
\auto\label{hf}
$$
where, as in Section 2, $m=1/2 + i\nu + n/2$ and
$\tilde{m}= 1/2 + i\nu -n/2$. We anticipate that, for the full kernel,
the eigenvalues are independent of $q^2$ in analogy with the $O(g^2)$
eigenvalues (because of the same dominance of large $k^2 \sim {k'}^2$). In this
case, the spectrum we have obtained already determines the holomorphic
factorization\cite{kir} and conformal symmetry properties of the non-forward
kernel.

We can rewrite (\ref{lam}) as
$$
\eqalign{ 16\pi\Lambda(\nu,n) ~=&~-~4\biggl(\beta'\bigl(m \bigr)
{}~+~\beta'\bigl(1-\tilde{m} \bigr)\biggr) \cr
=& ~\psi'\biggl({m+1 \over 2}\biggr) ~-~\psi'\biggl({m \over 2}\biggr)
{}~+~\psi'\biggl({2- \tilde{m} \over 2}\biggr)
{}~-~\psi'\biggl({1- \tilde{m} \over 2}\biggr) \cr
=& ~\sum_{r=0}^{\infty} {1 \over (r + {3 \over 4} + {n \over 4} +
{i\nu \over 2})^2}~-
{}~\sum_{r=0}^{\infty} {1 \over (r + {1 \over 4} + {n \over 4} +
{i\nu \over 2})^2}\cr
&+ ~\sum_{r=0}^{\infty} {1 \over (r + {3 \over 4} + {n \over 4} -
{i\nu \over 2})^2}
{}~-~ ~\sum_{r=0}^{\infty} {1 \over (r + {1 \over 4} + {n \over 4} -
{i\nu \over 2})^2} }
\auto\label{lam1}
$$
We next show that this expression is unchanged if we simultaneously send
$m \to 1-m$ and $ \tilde{m} \to 1-\tilde{m}$, i.e. $n \to -n, ~\nu \to -\nu$.
At this point it is crucial that $n$ is an even integer. Writing $n=2M$, we
obtain
$$
\eqalign{ 16\pi\Lambda(-\nu,-n) ~=&
\sum_{r=0}^{\infty} {1 \over (r + {1 \over 4} + {-M +1 \over 2} -
{i\nu \over 2})^2}~-
{}~\sum_{r=0}^{\infty} {1 \over (r + {1 \over 4} + {-M \over 2} -
{i\nu \over 2})^2}\cr
&+ ~\sum_{r=0}^{\infty} {1 \over (r + {1 \over 4} + {-M +1 \over 2} +
{i\nu \over 2})^2}
{}~-~ ~\sum_{r=0}^{\infty} {1 \over (r + {1 \over 4} + {-M \over 2} +
{i\nu \over 2})^2} }
\auto\label{lam2}
$$
and so
$$
\eqalign{ 16\pi\bigl(\Lambda(-\nu,-n) ~-~\Lambda(\nu,n)\bigr)&~=~
\sum_{s=-M}^{-1} {1 \over (s + {1 \over 4} + {M +1 \over 2} -
{i\nu \over 2})^2}~-
{}~\sum_{s=-M}^{-1} {1 \over (s + {1 \over 4} + {M \over 2} -
{i\nu \over 2})^2}\cr
+& ~\sum_{s=-M}^{-1} {1 \over (s + {1 \over 4} + {M +1 \over 2} +
{i\nu \over 2})^2}
{}~-~ ~\sum_{s=-M}^{-1} {1 \over (s + {1 \over 4} + {M \over 2} +
{i\nu \over 2})^2} \cr
&~=~\sum_{t=-M/2}^{M/2~-1} {1 \over (t + {3 \over 4} - {i\nu \over 2})^2}~-
{}~\sum_{t=-M/2}^{M/2~-1} {1 \over (-t  - {3 \over 4} - {i\nu \over 2})^2}\cr
+& ~\sum_{t=-M/2}^{M/2~-1} {1 \over (t + {3 \over 4} + {i\nu \over 2})^2}
{}~-~ ~\sum_{t=-M/2}^{M/2~-1} {1 \over (-t - {3\over 4} + {i\nu \over 2})^2}
\cr
&~=~~ 0 }
\auto\label{lam3}
$$

Because of this last symmetry, we can write
$$
\eqalign{ 16\pi\Lambda(\nu,n) ~=~-~2\biggl(\beta'\bigl(m \bigr) ~+~
\beta'\bigl(1-m \bigr)~+~\beta'\bigl(1-\tilde{m} \bigr)\biggr)
{}~+~\beta'\bigl(\tilde{m} \bigr)\biggr) }
\auto
$$
The final result needed to write (\ref{hf}) is that
$$
\eqalign{ \psi'\biggl({m \over 2}\biggr)
{}~+~\psi'\biggl({1-m \over 2}\biggr) ~&=~
\sum_{r=0}^{\infty}~{2r^2+r+1/4 -m(1-m)/2 \over
\bigl[r^2 + r/2 + m(1-m)/2 \bigr]^2}\cr
&=\tilde{{\cal F}_1}\bigl[m(1-m)\bigr] }
\auto
$$
and similarly
$$
\eqalign{ \psi'\biggl({m+1 \over 2}\biggr)
{}~+~\psi'\biggl({1-m+1 \over 2}\biggr) ~=~
\tilde{{\cal F}_2}\bigl[m(1-m)\bigr] ~.}
\auto
$$
This then allows us to write (\ref{hf}) with
$$
{\cal G}\bigl[m(1-m)\bigr]~=~{1 \over 8\pi}\biggl(
\tilde{{\cal F}_1}\bigl[m(1-m)\bigr]
{}~-~\tilde{{\cal F}_2}\bigl[m(1-m)\bigr] \biggr)~.
\auto
$$

Finally we discuss the numerical values that we obtain from our results.
The leading eigenvalue is at at $\nu=n=0$, as it is for the $O(g^2)$ kernel.
As we see from (\ref{gre1}), the correction to $\alpha_0$ is given by ${\cal
E}(0,0)/(16\pi^3)$. To evaluate this we use
$$
\eqalign{\Lambda(0,0) ~=& ~-~{1 \over 2\pi}\beta'(1/2)\cr
=& ~-~{1 \over 8\pi}\biggl(~\sum_{r=0}^{\infty} {1
\over (r+~1/4)^2} ~- ~\sum_{r=0}^{\infty} {1 \over (r+~3/4)^2}\biggr)\cr
&=~-~{1 \over 8\pi}\biggl(~16~+~{16 \over 25}~+~{16 \over 81}
{}~+~...~-~{16 \over 9}~-~{16 \over 49}~+~...\biggr)\cr
{}~&\sim ~-~ {1.81~ \over \pi}}
\auto
$$
{}From ${\cal K}_2$ alone we obtain
$$
\eqalign{ {9g^4\over 16\pi^3}\Lambda(0,0)
{}~\sim~-16.3 {{\alpha_s}^2 \over \pi^2} }
\auto
$$
The complete $\hat{K}^{(4n)}$ gives
$$
\eqalign{{\cal E}(0,0)/(16\pi^3)
{}~\sim&~{N^2g^4 \over 16 {\pi}^4}\biggl([2ln2]^2
{}~-~1.81 \biggr)\cr
{}~\sim&~{9g^4 \over 16{\pi}^4}\times 0.11 ~\sim~ {{\alpha_s}^2 \over
\pi^2} }
\auto
$$
giving a very small positive effect.

At this point we note that while ${\cal K}_1$ and ${\cal K}_2$ can
consistently be added to the leading-order kernel, this is not the case for
${\cal K}_0$. Although ${\cal K}_0$ is needed to regulate ${\cal
K}_1~+~{\cal K}_2$, it contains the $K^{(4)}_0$ diagrams which, as we
emphasized in \cite{uni}, can not be interpreted in terms of reggeization
effects. Since reggeization is the only consistent interpretation of
disconnected pieces, the $K^{(4)}_0$ diagrams can not be present in the full
kernel. As we described in \cite{spec}, to eliminate these diagrams while
retaining scale-invariance it is necessary to consider
$$
\eqalign{ \tilde{K}^{(4)}_{2,2} ~=~\hat{K}^{(4n)}_{2,2}
{}~-~\biggl(\hat{K}^{(2)}_{2,2}\biggr)^2~, }
\auto\label{k_4s}
$$
This is a consistent scale-invariant $O(g^4)$ kernel which can be added to
the $O(g^2)$ kernel. In this case, in writing (\ref{gre1}), we replace
${\cal E}(\nu,n)$ by $\tilde{{\cal E}}(\nu,n)$ where
$$
\tilde{{\cal E}}(\nu,n)~=~-~{3 \over \pi} [\chi(\nu,n)]^2~-~\Lambda(\nu,n) ~.
\auto
$$
This gives, as a modification of $\alpha_0$,
$$
\eqalign{{\tilde{{\cal E}}(0,0) \over 16\pi^3}
{}~=&~{N^2g^4 \over 16 {\pi}^4}\biggl(
-3 [\chi(0,0)]^2
{}~-~\Lambda(0,0) \biggr)\cr
\sim &~{9g^4 \over 16{\pi}^4}\times  (-5.76  - 1.81)\cr
\sim&~-68 {{\alpha_s}^2 \over \pi^2}}
\auto
$$
which is a substantial negative correction. We briefly discuss general
questions
related to the significance of these numbers in the next, concluding,
Section. They clearly show that, at next-to-leading-order,
the infra-red region can produce a strong reduction of the BFKL small-x
behavior.

\mainhead{8. CONCLUSIONS}

We have presented a variety of properties of the $O(g^4)$ kernel which appears
as an infra-red effect in the next-to-leading order BFKL equation. We have
emphasized the parallel between the mathematical properties of this new
kernel and the leading-order BFKL kernel. We have not addressed its general
significance. Indeed, in a recent paper Lipatov\cite{lip1} has argued that
the kernel obtained by direct next-to-leading order calculations will be
more complicated than our kernel and, in general, will not reduce to
transverse momentum integrals.

In a companion paper \cite{exp} we show that the kernels we have studied
unambigously arise when the $t$-channel unitarity equations are, in a
weak-coupling approximation, systematically expanded in the angular momentum
plane around $j~=~1$, while keeping the leading particle singularities.
Consequently the kernels we have studied must be obtainable as infra-red (in
transverse momentum) approximations to the full next-to-leading log kernel
calculated in momentum space\cite{lip1}. We find that $K^{4n}$ enters
directly as a next-to-leading order scale-invariant contribution.
$\bigl(K^{(2)}\bigr)^2$ also appears at next-to-leading-order but,
simultaneously, internal transverse momentum logarithms appear which violate
the scale-invariance. These are the logarithms found by Fadin and
Lipatov\cite{fad} in the higher-order trajectory function. Therefore it is
not clear that is sensible to maintain scale-invariance when introducing
$\bigl(K^{(2)}\bigr)^2$.

Since the $t$-channel unitarity equations are infra-red constraints,
the use of transverse-momentum integrals may only be valid below some
transverse momentum cut-off. For applications to small-x physics at
relatively large transverse momentum, this cut-off could be crucial for a
reliable estimate of the relative magnitude of our contributions in the full
physical kernel. Because of this, the numerical results we obtain tell us only
the size of effects that are obtained by naively extending the infra-red
behaviour of the theory to arbitrarily large transverse momentum. In this
context we should remark that it is not yet clear that the program of
\cite{lip1} leads to a next-to-leading order kernel that can be
consistently evaluated at large transverse momentum - with the running
coupling properly involved.

Ultimately it is the Regge limit of QCD which is of most theoretical interest.
The Regge region is truly the infra-red region in transverse momentum and this
is what $t$-channel unitarity is best suited to studying. We argue in our
companion paper that the expansion techniques we develop, when properly
limited to the infra-red region, are likely to be very powerful in providing
very high-order information on the Regge limit. Indeed, we believe the
scale-invariant kernels we derive may play a fundamental role in
understanding the true nature of the soft Pomeron as a full solution of QCD
at asymptotic energies.

It is from this last perspective that it seems most important to study to what
extent fundamental properties of the leading-order kernel, such as holomorphic
factorization and conformal symmetry, are preserved in higher-order
scale-invariant kernels. That the eigenvalue spectrum of the $O(g^4)$ kernel
does indeed have such properties, seems to us both interesting and
intriguing.

\vspace{1cm}
\centerline{\bf Acknowledgements}
We thank J. Bartels, V. Fadin, R. Kirschner and L. Lipatov for informative
discussions and comments. C.C. is grateful to B. Andersson,
Yu. L. Dokshitzer and to G. Marchesini for very formative lectures and
discussions at Lund Univ. and at St. Petersburg's winter school.

\vspace{.2in}

\renewcommand{\theequation}{A.\arabic{equation}}
\setcounter{equation}{0}
\noindent {\large\bf Appendix A. Kinematics }
\vskip 3mm
\noindent We review here some kinematic features which are special to two
dimensions and which complicate our analysis.

In two dimensions there is only one independent two-particle invariant in
elastic scattering. That is both $t$ and $u$ can be choosen to be functions
of $s$ which, as usual we interpret as the total energy. In fact, consider a
2-to-2 particle scattering $k_1+k_2~\to~ k_3 + k_4$. If we define
$$
z_i~~=~~k_i^2~~~~i~=~1,..,4 ,
$$
then in the center of mass we have

\beqa
&& k_1={1\over 2 \sqrt{s}}\left(s - {k_2}^2+ {k_1}^2,\lambda^{1/2}(
s,k_1^2,k_2^2)\right)\nonumber \\
&& k_2={1\over 2 \sqrt{s}}\left(s + {k_2}^2 - {k_1}^2,-\lambda^{1/2}(
s,k_1^2,k_2^2)\right)\nonumber \\
&& k_3={1\over 2 \sqrt{s}}\left(s +k_3^2 - k_4^2,\lambda^{1/2}(
s,k_3^2,k_4^2)\right)\nonumber \\
&& k_4={1\over 2 \sqrt{s}}\left(s -k_3^2 + k_4^2,-\lambda^{1/2}(
s,k_3^2,k_4^2)\right).\nonumber \\
\label{center}
\eeqa
One still defines the Mandelstam invariants as in D=4:
 $s$, $t=(k_3-k_1)^2$ and $u=(k_1-k_4)^2$, and verifies that they satisfy
the usual Mandelstam relation
\beqa
s+t+u=(k_1^2+k_2^2+k_3^2+k_4^2)~.
\eeqa
{}From (\ref{center}), however, it is easy
to derive the relations
\beqa
&& t={k_1^2+k_2^2 +k_3^2+k_4^2 -s\over 2}
-{(-k_2^2+k_1^2)(-k_4^2 +k_3^2)-
\lambda^{1/2}(s,k_1^2,k_2^2)\lambda^{1/2}(s,k_3^2,k_4^2)
\over 2 s} \nonumber \\
&& u={k_1^2+ k_2^2+k_3^2+k_4^2 -s\over 2}
-{(-k_2^2 +k_1^2)(k_4^2-k_3^2)+
\lambda^{1/2}(s,k_1^2,k_2^2)\lambda^{1/2}(s,k_3^2,k_4^2)
\over 2 s}. \nonumber \\
\label{stu}
\eeqa

\renewcommand{\theequation}{B.\arabic{equation}}
\setcounter{equation}{0}
\noindent {\large\bf Appendix B. Evaluation of the coefficients $A_{ij}$ }
\vskip 3mm
\noindent In this Appendix we illustrate in some more detail the evaluation
of the coefficients of the logarithms of the box diagram. As we emphasized
in the text, it is crucial to use explicit expressions for the dual vectors
${k^d}_i$ in terms of the original external momenta $k_i$ in order to obtain
a simple result for the $A_{ij}$'s. We first evaluate the $A^\pm_{ij}$ with a
mass cut-off included.

Noting that from (\ref{prod})
\beq
A^{\pm}_{12}={1\over ((q^{\pm}_{12}-p_3)^2 + m^2)((q^{\pm}_{12}-p_4)^2+ m^2)}
\eeq
we use the definition of $q^{\pm}_{12}$ in (\ref{dua}) to obtain
\beqa
&& ((q^{\pm}_{12}-p_3)^2 + m^2)((q^{\pm}_{12}-p_4)^2+ m^2)\nonumber \\
&&={ 1\over n_{12}z_1^2}\left(z_1 z_2 + z_1 k_1\cdot k_2\pm n_{12}
\lambda^{1/2}(-z_1,m^2,m^2)\right)\nonumber \\
&&\,\,\,\,\,\,\,\times\left( z_1 (z_2 + z_3 -2 k_2\cdot k_3 + k_1\cdot k_2 -
k_1\cdot k_3) n_{12}\right. \nonumber \\
&&\left. \pm
((k_1\cdot k_2)^2 - z_1 z_2 - k_1\cdot k_3 k_1\cdot k_2 +
z_1 k_2\cdot k_3)\lambda^{1/2}(z_1,m^2,m^2)\right)
\eeqa
where
\beq
z_i=k_i^2,\,\,\,\,\,\, n_{ij}=\sqrt{(k_i\cdot k_j)^2- z_i z_j}.
\eeq
Similarly
\beq
A^{\pm}_{13}={1\over ((q^{\pm}_{13}-p_2)^2 + m^2)(q^{\pm}_{13}-p_4)^2 + m^2)}
\eeq
with
\beqa
&&((q^{\pm}_{13}-p_2)^2 + m^2)(q^{\pm}_{13}-p_4)^2 + m^2)\nonumber \\
&&=- {k_1\cdot k_2\over s}
\left( s\,\,\, n_{12}(z_3 - k_1\cdot k_3 - k_2\cdot k_3)\right.\nonumber \\
&&\,\,\,\,\,\,\left.\pm \lambda^{1/2}(s, m^2,m^2)(k_1\cdot k_3 k_1\cdot k_2
+k_2\cdot k_3 k_1\cdot k_2 - z_1 k_2\cdot k_3 - z_2 k_1\cdot k_3)\right)
\nonumber
\eeqa
\beq
A^{\pm}_{14}={1\over ((q^{\pm}_{14}-p_2)^2 + m^2)(q^{\pm}_{14}-p_3)^2 + m^2)}
\eeq
with
\beqa
&&((q^{\pm}_{14}-p_2)^2 + m^2)((q^{\pm}_{14}-p_3)^2 + m^2)\nonumber \\
&& ={1\over n_{34} k_4^2}
\left( z_4 (z_3 + z_2 -2 k_2\cdot k_3 +k_3\cdot k_4 - k_2\cdot k_4) n_{34}
\right. \nonumber \\
&& \,\,\,\,\,\,\,\,\left. \pm (k_3\cdot k_4 k_2\cdot k_4 - k_3\cdot k_4
k_3\cdot k_4 -
z_4 (k_2\cdot k_3 - z_2))\lambda^{1/2}(k_4^2,m^2,m^2)\right) \nonumber \\
&&\,\,\,\,\,\,\times \left(z_3 z_4 + z_4 k_3\cdot k_4 \pm n_{34}\lambda^{1/2}
(-k_4^2,m^2,m^2)\right) \nonumber \\
\eeqa
\beq
A^{\pm}_{23}={1\over ((q^{\pm}_{23}-p_4)^2 + m^2)(q^{\pm}_{23}-p_1)^2 + m^2)}
\eeq
with
\beqa
&&((q^{\pm}_{23}-p_4)^2 + m^2)((q^{\pm}_{23}-p_1)^2 + m^2)\nonumber \\
&&= {1\over n_{12}z_2^2}\left( z_1 z_2 + z_2 k_1\cdot k_2
\mp n_{12}\lambda^{1/2}(z_2,m^2,m^2)\right)\nonumber \\
&&\,\,\,\,\,\,\,\,\,\times \left(
(z_2 z_3 - z_2 k_2\cdot k_3)n_{12} \mp (k_2\cdot k_3 k_1\cdot k_2 -
z_2 k_1\cdot k_3)\lambda^{1/2}(z_2,m^2,m^2)\right) \nonumber  \\
\eeqa
\beq
A^{\pm}_{24}={1\over ((q^{\pm}_{24}-p_1)^2 + m^2)(q^{\pm}_{24}-p_3)^2 + m^2)}
\eeq
with
\beqa
&&((q^{\pm}_{24}-p_1)^2 + m^2)((q^{\pm}_{24}-p_3)^2 + m^2)\nonumber \\
&& ={1\over (k_1-k_3)^2 n_{12}^2 n_{34}}
\left(n_{12}n_{34}(z_1 - k_1\cdot k_3 + k_1\cdot k_2)(k_2-k_3)^2
\pm H_{24,1}\right)\nonumber \\
&& \,\,\,\,\,\,\left( k_2\cdot k_3 (k_2-k_3)^2 n_{12}\pm H_{24,3}\right)
\nonumber \\
\eeqa
where
\beqa
&& H_{24,1}=\left( (k_1\cdot k_2)^2 - z_1 z_2\right) n_{34}
 -\left(k_1\cdot k_3  k_3\cdot k_4 - z_3 k_1\cdot k_4\right)
\lambda^{1/2}((k_2-k_3)^2,m^2,m^2)\nonumber \\
&& H_{24,3}=\left( k_2\cdot k_3 k_1\cdot k_2 - z_2 k_1\cdot k_3\right)
\lambda^{1/2}((k_2-k_3)^2,m^2,m^2)\nonumber \\
\eeqa
Finally
\beq
A^{\pm}_{34}={1\over ((q^{\pm}_{34}-p_1)^2 + m^2)(q^{\pm}_{34}-p_2)^2 + m^2)}
\eeq
with
\beqa
&&((q^{\pm}_{34}-p_1)^2 + m^2)((q^{\pm}_{34}-p_2)^2 + m^2)\nonumber \\
&&={1\over z_3^2 n_{34}}\left( z_1 + z_2 +2 k_1\cdot k_2 -
k_1\cdot k_3 -k_2\cdot k_3 \mp n_{34}\lambda^{1/2}(k_3^2,m^2,m^2)\right)
\nonumber\\
&&\times \,\,\,\,\,\, \left((z_2 z_3 - z_3 k_2\cdot k_3)n_{34}\mp H_{34,2}
\lambda^{1/2}(k_3^2,m^2,m^2)\right)
\eeqa
where
\beq
H_{34,2}=k_2\cdot k_3 k_3\cdot k_4 - z_2 k_2\cdot k_4.
\eeq
Notice that all the coefficients $ A_{ij}$ can be re-expressed in terms of
the invariant $s=(k_1+k_2)^2$ quite straightforwardly.

It is because each dual vector contains non-integer powers of the
$\Lambda$-function that it is useful to combine the single contributions
$A^\pm_{ij}$ with a single common denominator, to obtain a simple polynomial
result. Combining this with the massless limit leads to the
$A_{ij}~\equiv~a_{ij}/b_{ij}$ utilised in the text. The $a_{ij}$ and $b_{ij}$
that we did not give explicitly in Section 4 are
\beqa
&&b_{34} =\biggl[ -{{\left( - {  k_2} \cdot {  k_4}\,{  k_3}^{2}   +
         {  k_2} \cdot {  k_3}\,{  k_3} \cdot {  k_4} \right) }^2} +
   {{( {  k_2}^{2} - {  k_2} \cdot {  k_3} ) }^2}\,
    ( {{{  k_3} \cdot {  k_4}}^2} - {  k_3}^{2}\,{  k_4}^{2} ) \biggr]
\nonumber \\
&&\times\biggl[ -{{\left( - ( {  k_1} \cdot {  k_4} +
              {  k_2} \cdot {  k_4} ) \,{  k_3}^{2}   +
         ( {  k_1} \cdot {  k_3} + {  k_2} \cdot {  k_3}
            ) \,{  k_3} \cdot {  k_4} \right) }^2}\biggr.\nonumber \\
&&\,\,\,\,\,\,\,\,\,\biggl. +
   {{( {  k_1}^{2} + 2\,{  k_1} \cdot {  k_2} -
         {  k_1} \cdot {  k_3} + {  k_2}^{2} - {  k_2} \cdot {  k_3}\
          ) }^2}\,( {{{  k_3} \cdot {  k_4}}^2} -
      {  k_3}^{2}\,{  k_4}^{2} ) \biggr] \nonumber
\eeqa
\beqa
&&a_{14} =\biggl[ {{{  k_3} \cdot {  k_4}}^2} - {  k_3}^{2}\,{  k_4}^{2}
 \biggr]\nonumber \\
&&\,\,\,\,\,\,\times\biggl[ {  k_3} \cdot {  k_4}\,
    ( -{  k_2} \cdot {  k_4} + {  k_3} \cdot {  k_4} )\biggr. \nonumber \\
&&\biggl. +
   ( {  k_3}^{2} + {  k_3} \cdot {  k_4} ) \,
    ( {  k_2}^{2} - 2\,{  k_2} \cdot {  k_3}
 - {  k_2} \cdot {  k_4} + {  k_3}^{2} + {  k_3} \cdot {  k_4})
    - ( {  k_3}^{2} - {  k_3} \cdot {  k_4} ) \,
    {  k_4}^{2}\biggr] \nonumber \\
\eeqa
\beqa
&&b_{14} =\biggl[ -{{{  k_3} \cdot {  k_4}}^2} +
   {{( {  k_3}^{2} + {  k_3} \cdot {  k_4} ) }^2} +
   {  k_3}^{2}\,{  k_4}^{2}\biggr]\nonumber \\
&&\,\,\,\,\,\,\,\,\,\,\times\biggl[{{( {  k_2}^{2} - 2\,{  k_2} \cdot {  k_3} -
         {  k_2} \cdot {  k_4} + {  k_3}^{2} + {  k_3} \cdot {  k_4}
          ) }^2}\,( {{{  k_3} \cdot {  k_4}}^2}-
      {  k_3}^{2}\,{  k_4}^{2} ) \biggr.\nonumber \\
&&\biggl. -
   {{[ {  k_3} \cdot {  k_4}\,
         ( -{  k_2} \cdot {  k_4} + {  k_3} \cdot {  k_4}
           )  - ( {  k_3}^{2} - {  k_3} \cdot {  k_4} ) \,
         {  k_4}^{2} ] }^2}\biggr] \nonumber \\
\eeqa
\beqa
&&a_{13} =\biggl[ {{{  k_1} \cdot {  k_2}}^2} - {  k_1}^{2}\,{  k_2}^{2}\biggr]
\nonumber \\
&&\,\,\,\,\,\,\times\biggl[ {  k_1} \cdot {  k_2}\,{  k_1} \cdot {  k_3} +
   {  k_1} \cdot {  k_3}\,{  k_2}^{2} -
   {  k_1}^{2}\,{  k_2} \cdot {  k_3} -
   {  k_1} \cdot {  k_2}\,{  k_2} \cdot {  k_3} +
   {  k_1} \cdot {  k_2}\,{  k_3} \cdot {  k_4}\biggr]\nonumber \\
\eeqa
\beqa
&&b_{13} = {  k_1}^{2} {  k_2}^{2}
\left[ -{{( {  k_1} \cdot {  k_2}\,{  k_1} \cdot {  k_3} +
         {  k_1} \cdot {  k_3}\,{  k_2}^{2} -
         {  k_1}^{2}\,{  k_2} \cdot {  k_3} -
         {  k_1} \cdot {  k_2}\,{  k_2} \cdot {  k_3} ) }^2}\right.\nonumber \\
&&\left. +
   ( {{{  k_1} \cdot {  k_2}}^2} - {  k_1}^{2}\,{  k_2}^{2}
      ) \,{{{  k_3} \cdot {  k_4}}^2}\right]\nonumber \\
\eeqa
\beqa
&&a_{24} =\biggl[ {  k_1} \cdot {  k_3}\,{  k_2}^{2} -
   {  k_1} \cdot {  k_2}\,{  k_2} \cdot {  k_3} +
   ( {  k_1}^{2} + {  k_1} \cdot {  k_2} \nonumber \\
&&\,\,\,\,\,\,\,\,\,\,- {  k_1} \cdot {  k_3} ) \,{  k_2} \cdot {  k_3} -
   {  k_1} \cdot {  k_3}\,{  k_2} \cdot {  k_3} +
   {  k_1} \cdot {  k_2}\,{  k_3}^{2}\biggr]
\biggl[ {{{  k_2} \cdot {  k_3}}^2} - {  k_2}^{2}\,{  k_3}^{2}\biggr]
\nonumber \\
\eeqa
\beqa
&&b_{24}={  k_2}^{2} {  k_3}^{2}
\biggl[  -{{( {  k_1} \cdot {  k_3}\,{  k_2}^{2} -
         {  k_1} \cdot {  k_2}\,{  k_2} \cdot {  k_3} -
         {  k_1} \cdot {  k_3}\,{  k_2} \cdot {  k_3} +
         {  k_1} \cdot {  k_2}\,{  k_3}^{2} ) }^2} \biggr.
\nonumber \\
&&\biggl.+
   {{( {  k_1}^{2} + {  k_1} \cdot {  k_2} -
         {  k_1} \cdot {  k_3} ) }^2}\,
    ( {{{  k_2} \cdot {  k_3}}^2} - {  k_2}^{2}\,{  k_3}^{2} ) \biggr]
\eeqa

\renewcommand{\theequation}{C.\arabic{equation}}
\setcounter{equation}{0}
\noindent {\large\bf Appendix C. Dual Vectors }
\vskip 3mm
\noindent The definitions of the dual vectors $k^d_i$  eq.~(\ref{duals})  in
terms of the external momenta $k_i$ are not unique. For instance we could
have defined
\beqa
&& {k^d}_1={\epsilon(n_{13})\over \sqrt{(k_1\cdot k_3)^2 - k_1^2 k_3^2}}
\left( k_1 k_1\cdot k_3 - k_3 k_1^2\right) \nonumber \\
&& {k^d}_2=-{\epsilon(n_{23})\over \sqrt{(k_2\cdot k_3)^2 - k_2^2 k_3^2}}
\left( k_2 k_2\cdot k_3 - k_3 k_2^2\right), \nonumber \\
\label{duals1}
\eeqa
and similar others. It is also easily shown that
\beqa
&& \lambda(t, k_1^2,k_3^2)=\epsilon(n_{13})
 ((k_1\cdot k_3)^2 - k_1^2 k_3^2) \nonumber \\
&& \lambda(u, k_1^2,k_4^2)=\epsilon(n_{14})
((k_1\cdot k_4)^2 - k_1^2 k_4^2). \nonumber \\
\eeqa
Nontrivial relations between $\Lambda$-functions can also be derived e.g.
\beqa
 \lambda^{1/2}(s,k_1^2,k_2^2)\lambda^{1/2}(t,k_1^2,k_3^2)
&=&4 ( k_1\cdot k_2 k_2\cdot k_3 - k_2^2 k_1\cdot k_3)
\eeqa
and
\beqa
{\lambda^{1/2}(t,k_2^2,k_3^2)\over \lambda(s,k_1^2,k_2^2)}
&=&{\epsilon(n_{23})(k_1\cdot k_3 k_2\cdot k_3 - k_3^2 k_1\cdot k_2)\over
\epsilon(n_{12})(k_1\cdot k_3 k_1\cdot k_2
- k_1^2 k_2\cdot k_3)}
\label{lambdas}
\eeqa
(valid for positive values of the left-hand sides) in the Minkowsky region.
In the euclidean region, exploiting the $\lambda$-function as an area
(see Fig.~C.1) gives
\beqa
&& \lambda^{1/2}(s, k_1^2,k_2^2) + \lambda^{1/2}(s, k_3^2,k_4^2)=
\lambda^{1/2}(t, k_1^2,k_3^2) +\lambda^{1/2}(t, k_2^2,k_4^2).
\eeqa
The existence of these relations is what makes simplification of the
coefficients $A_{ij}$ non-trivial.

\vspace{.2in}

\epsffile{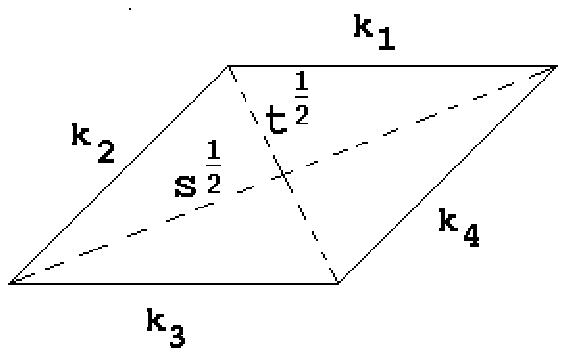}
\begin{center}
Fig.~C.1 Euclidean triangles
\end{center}

\renewcommand{\theequation}{D.\arabic{equation}}
\setcounter{equation}{0}
\vskip 1cm \noindent
\noindent {\large\bf Appendix D. Analyticity Properties.}
\vskip 3mm \noindent In this Appendix we discuss some aspects of the
analyticity properties and dispersion theory of one-loop n-point functions
at $D=2$. Note first that in the special case in which any of the external
momenta becomes exceptional, the Kallen-Toll method is not applicable.
Nevertheless it is still possible to evaluate directly the box diagram
integral, for example, from its dispersive part, i.e. we write
\beq
I(k_1,k_2,k_3,k_4)={1\over 4 \pi}
\int_{4 m^2}^{\infty} dz{\Delta(z)\over (z-s)}
\eeq

where
\beq
\Delta(z)={\sqrt{z- 4 m^2}\lambda[z, k_1^2,k_2^2]^{1/2}
\lambda[s, k_3^2, k_4^2]^{1/2}\over (4 m^2 \lambda[z, k_1^2,k_2^2] + 4
z k_1^2 k_2^2)(4 m^2 \lambda[z, k_3^2,k_4^2] + 4 z k_3^2 k_4^2)}
\eeq
is the $s-$channel cut of the box diagram
\beq
\Delta(s)=\int\,d^4\,q {\delta_+(q^2-m^2)\delta_+((k_1+k_2-q)^2- m^2)\over
[(k_1-q)^2 - m^2][k_2-q)^2-m^2]}
\eeq

The dispersion integral is not straightforward, in the general case,
and involves also elliptic functions. The logarithmic structure may appear only
after non-trivial manipulations. However, in the case of exceptional
external momenta it provides a simple alternative to the method of residui.
We remark that all the one-loop functions can be re-obtained by dispersive
methods quite elementarily. This is particularly simple when, as in our case,
the internal masses are equal and so there is no contribution from
anomalous thresholds.

We also observe that the 2-point function
\beq
I(q^2)~=~=~16\pi^3J_1(q^2,m^2)~=~\int {dp\over (p^2+ m^2)((q-p)^2 + m^2}
\eeq
can be written down in various forms. These forms may differ by the choice of
phase conventions for the logarithms involved. For instance one easily
obtains
\beq
I(q)={\theta\over m^2 sinh\,\, \theta}~~~~~~~~~~~~~~~~~~~~~
cosh\,\,\theta={ q^2 + 2 m^2\over 2 m^2}
\eeq
which in its logarithmic form becomes
\beq
I(q)={ 2\over \sqrt{q^2(q^2 + 4 m^2)}}Log\left({
 q^2 + 2 m^2 \pm \sqrt{q^2 + 4 m^2}\over 2 m^2}\right).
\label{loga}
\eeq
If we choose either the ``~+~'' or ``~-~'' determination of the logarithm
in (\ref{loga}), we encounter a branch cut at $q^2=-4 m^2$ and $no$
threshold singularity (at $q=0$), as expected. Combinining both
determinations, instead, we obtain the Kallen-Toll result for $I(q)$ quoted
in Section 4 which is free of any singularity in the finite plane. In our case
the distinction is not essential, since we are only interested in real parts
obtained in the massless limit.

\vspace{.1in}
\renewcommand{\theequation}{E.\arabic{equation}}
\setcounter{equation}{0}
\noindent {\large\bf Appendix E. Identities. }

\vspace{.1in}

\noindent In Dimensional Regularization we get
\beq
{1\over [k^2]^\alpha}={\Gamma[D/2-\alpha]\over \Gamma[\alpha]\pi^{D/2}}
\int {d^Dx e^{2 i k\cdot x}\over [x^2]^{D/2-\alpha}}.
\eeq
Using this relation we find a simple expression for the integral
\beqa
 I[R,S]~&=&~\int {d^D k\over{[(k-q)^2]^R [k^2]^S}} \nonumber \\
 &=&~ {\Gamma[D/2-R]\Gamma[D/2-S]\over (2\pi)^D\Gamma[R]\Gamma[S]}
\int {d^Dx e^{2 i q\cdot x}\over [x^2]^{D/2-\sigma}}\nonumber \\,
\eeqa
where $\sigma=R +S- D/2$. A simple manipulation of this expression gives
\beq
I[R,S]={1\over [q^2]^{R+S-D/2}}
{\Gamma[D/2-R]\Gamma[D/2-S]\Gamma[R+S-D/2]\over
\pi^{D/2}2^D \Gamma[R]\Gamma[S]\Gamma[D-R-S]}
\eeq

$J_1(k^2)$ is immediately evaluated as in the text. For $J_2(k^2)$ we obtain
\beqa
&& J_2(k^2)\equiv\int d^D p {J_1(p)\over (k-p)^2} \nonumber \\
&& \,\,\,\,\,={1\over [k^2]^{3-D}}
{\eta\ \pi^{D/2}\Gamma[D/2-1]\Gamma[D-2]\Gamma[3-D]\over
\Gamma[2-D/2]\Gamma[3 D/2 -3]}.
\eeqa
Writing $D=2+\epsilon$ we obtain
\beqa
&&k^2J_2(k^2) = {{6 {{{\rm \gamma}}^2} {{\pi }^2}}} -
   {{{{\pi }^4}}\over {2 }} + {{12 {{\pi }^2}}\over {{{{\it \epsilon}}^2} }} +
   {{12 {\rm \gamma} {{\pi }^2}}\over {{\it \epsilon} }} +
   {{12 {\rm \gamma} {{\pi }^2} \log (\pi )}} +
   {{12 {{\pi }^2} \log (\pi )}\over {{\it \epsilon} }} +
   {{6 {{\pi }^2} [\log (\pi )]^2}} \nonumber \\
&& +
   {{12 {\rm \gamma} {{\pi }^2} \log (k^2)}} +
   {{12 {{\pi }^2} \log (k^2)}\over {{\it \epsilon} }}
+
   {{12 {{\pi }^2} \log (\pi ) \log (k^2)}} +
   {{6 {{\pi }^2} [\log (k^2]^2}}.\nonumber \\
\eeqa
Similarly we obtain
\beqa
&&  \left(k^2J_1(k)\right)^2 = 8 {{{\rm \gamma}}^2} {{\pi }^2}  -
   {{2 {{\pi }^4} }\over 3} +
   {{16 {{\pi }^2} }\over {{{{\it \epsilon}}^2}}}
+ {{16 {\rm \gamma} {{\pi }^2} }\over {{\it \epsilon}}} +
   16 {\rm \gamma} {{\pi }^2} \log (\pi ) +
   {{16 {{\pi }^2}  \log (\pi )}\over {{\it \epsilon}}} +
   8 {{\pi }^2}  [\log (\pi )]^2 \nonumber \\
&&\left. + 16 {\rm \gamma} {{\pi }^2}  \log (k^2) +
   {{16 {{\pi }^2} \log (q^2)}\over {{\it \epsilon}}} +
   16 {{\pi }^2} \log (\pi ) \log (k^2) +
   8 {{\pi }^2}  [\log (k^2)]^2\right., \nonumber \\
\eeqa

We can also extract the leading double logarithmic contributions by
introducing a mass cutoff. That is we write
\beq
J_2(k^2)\to J_2(k^2,m^2)=\int {d^2q\over [(k-q)^2 + m^2]}
\int {d^2l\over [l^2 +m^2][(l-q)^2 + m^2]}
\eeq
Perfoming the $l$ integral and one angular integral, and dropping
the cutoff dependence whenever possible we get
\beqa
&& J_2(k^2,m^2)=-\pi\int_{0}^{\infty}{d\,q^2\over (k^2 + q^2)
\sqrt{q^2 (q^2 + 4 m^2)}}Log[\chi(q^2)],\nonumber \\
&& \chi(q^2)={1- \sqrt{1- 4 m^2/q^2}\over 1 + \sqrt{1 - 4 m^2/q^2}}.
\nonumber \\
\eeqa
Defining the change of variable
\beqa
&& x=\chi(q^2), \,\,\,\,if\,\,\,\,\,  q^2\neq 0\nonumber \\
&& x=-1\,\,\,\,\,\, if\,\,\,\,\,\, q^2=0 \nonumber \\
\eeqa
some manipulations allow us to rewrite the integral in the form
\beq
J_2(k^2,m^2)=\pi\int_{1}^{\infty}d\,\,x\,\,{Log[x]\over (x + x_0)(x + x'_0)}
\label{logo}
\eeq
with
\beq
 x_0={1\over x'_0}={1+ \sqrt{1+ 4 m^2/q^2}\over 1 - \sqrt{1 + 4 m^2/q^2}}.
\eeq
The last integration in (\ref{logo}) can be easily performed
by using
$$
\int_0^{z}dx \,\,{Log(x-a)\over x-b}=~
\Biggl[~Sp\biggl[{a-b\over x-b}
\biggr] + {1\over 2}Log[x-b]^2~\Biggr]^z_1 \nonumber \\
\auto
$$
to obtain
\beq
J_2(k^2,m^2)={\pi\over 2 (x'_0-x_0)}\left(Log^2[1+x'_0] -Log^2[1 + x_0]\right)
\eeq
which clearly exhibits the $Log^2[m^2/k^2]$ contributions.

\vspace{.1in}
\renewcommand{\theequation}{F.\arabic{equation}}
\setcounter{equation}{0}
\noindent {\large\bf Appendix F. Impact Parameter Space. }

\vspace{.1in}

\noindent We note first that the contours in (\ref{mell}) separate the right
and left poles of the $\Gamma$ functions. The contour can be closed in
various possible ways, thus generating ascending and descending series, as
usually happens for the Mellin transform. To derive the Gegenbauer expansion
of the {\cal Y}, at some special value of the powers of the propagators, we
proceed as follows.

Following ref. \cite{chet} we start by defining
\beq
S_\lambda\equiv{ 2 \pi^{\lambda + 1}\over \Gamma[\lambda +1]}
\eeq
and introduce the expansion
\beq
{1\over [(p-q)^2]^\lambda}={1\over {d_>}^{2 \lambda} (p,q)}\sum_{n=0}^{\infty}
{d_<}^n(p,q) C_n^{\lambda}(\hat{p}\cdot \hat{q}).
\eeq
We have defined
\beqa
&& d_<(p,q)=min({p\over q},{q\over p})\nonumber \\
&& d_>(p,q)=Max(p,q). \nonumber \\
\eeqa

The $D$ dimensional integration measures are then re-expressed as
$d^D z=z^{2\lambda +1}S_\lambda d\hat{z}$ which isolates the $\hat{ z}$
part of the integration. By using the orthogonality relation
\beq
\int d\hat{p} C_{n_1}^\lambda(\hat{k}\cdot \hat{p})C_{n_2}^{\lambda}
(\hat{p}\cdot {q})=\delta_{n_1\, n_2}{\lambda\over n_1 +
\lambda}C_{n_1}^\lambda
(\hat{k}\cdot \hat{q})
\eeq
it is easy to derive the expression for $Y(\sigma_1,D/2-1,D/2-1|p,q)$ as a
Gegenbauer series. The expression of ${\cal J}_n(x_+,x_-)$ is given by
\beqa
&& {\cal J}_n(x_+,x_-)=\theta(x_+-x_-) y(x_+,x_-) + \theta(x_--x_+) y(x_-,x_+),
\eeqa
with
\beqa
&&y(x_+,x_-)={x_-^{2 + n - 2 \sigma_1} x_+^{2 - D - n}\over D + 2 n -2
\sigma_1}
\nonumber \\
&& + {x_+^{4 -2\sigma_1 -D-n}x_-^{n }\over 2(1-\sigma_1)}
-{x_-^{2 - 2 \sigma_1 +n}x_+^{2 -D-n}\over 2 (1-\sigma_1)}\nonumber \\
&&- {x_-^n x_+^{4- n -2\sigma_1 -D}\over 4 -2 n-2 \sigma_1 -D}\nonumber \\
\eeqa

\end{document}